\documentclass{article}
\usepackage[utf8]{inputenc}
\usepackage[T1]{fontenc}
\usepackage{arxiv}
\usepackage{bm}
\usepackage{amsmath,amssymb}
\usepackage{graphicx}
\usepackage{booktabs}
\usepackage{hyperref}
\usepackage{url}
\usepackage{color}
\usepackage{tikz}
\usepackage{tikz,amsmath,amssymb}
\usetikzlibrary{arrows.meta,positioning,calc}
\usepackage{pgfplots}
\pgfplotsset{compat=1.18}
\usepackage{subcaption}
\usetikzlibrary{decorations.pathreplacing,calc,positioning}
\usetikzlibrary{patterns}
\usepackage{multirow}

\title{Stable Audio 3}

\author{
  Zach Evans, Julian D. Parker, Matthew Rice, CJ Carr, Zack Zukowski, Josiah Taylor, Jordi Pons
}

\date{}

\begin{document}

\maketitle

\vspace{-6mm}

\begin{abstract}

Stable Audio 3 is a family of fast latent diffusion models (\texttt{small}, \texttt{medium}, \texttt{large}) for variable-length audio generation and editing. Since our models can generate several minutes of audio, variable-length generations are key to avoid the cost of producing full-length generations for short sounds. We also support inpainting, enabling targeted audio editing and the continuation of short recordings. Our latent diffusion models operate on top of a novel semantic-acoustic autoencoder that projects audio into a compact latent space, enabling efficient diffusion-based generation while preserving audio fidelity and encouraging semantic structure in the latent. 
Finally, we run adversarial post-training to both accelerate inference and improve generation quality, reducing the number of inference steps while improving fidelity and prompt adherence.
Stable Audio 3 models are trained on licensed and Creative Commons data to generate music and sounds in less than a 2s on an H200 GPU and less than a few seconds on a MacBook Pro M4.
We release the weights of \texttt{small} and \texttt{medium}, that can run on consumer-grade hardware, together with their training and inference pipeline.
\begin{center}
    \url{https://github.com/Stability-AI/stable-audio-tools}
    \url{http://github.com/Stability-AI/stable-audio-3}    
\end{center}

\end{abstract}

\begin{figure}[h]
\centering
\begin{tikzpicture}[
    >=stealth,
    block/.style={rectangle, draw, rounded corners=4pt,
        minimum height=0.6cm, inner sep=5pt,
        font=\footnotesize, align=center},
    wave/.style={font=\footnotesize, align=center, inner sep=2pt},
    arr/.style={->, semithick},
    cond/.style={font=\footnotesize, align=center},
]

\tikzset{miniwave/.style={
    baseline=-0.5ex,
    scale=0.12,
    line width=0.5pt
}}

\node[wave] (wav_in) {
\begin{tikzpicture}[miniwave]

\draw[blue!70!black] plot[domain=0:12.56,samples=160]
(\x,{0.6*sin(2*\x r)});

\draw[red!70!black] plot[domain=0:12.56,samples=160]
(\x,{0.6*sin(2*(\x-0.4) r)});

\end{tikzpicture}\\[2pt]
{\scriptsize\color{gray} stereo 44.1\,kHz}
};

\node[block, right=0.6cm of wav_in] (enc) {semantic-acoustic\\encoder\\{\scriptsize\color{gray}4096$\times$ downsampling}};

\node[block, fill=gray!12, right=0.6cm of enc,
      minimum width=5.2cm, minimum height=1.0cm] (dit)
{latent diffusion transformer\\{\scriptsize\color{gray}variable length inference}\\{\scriptsize\color{gray}long-form and fast generation}};

\node[block, right=0.6cm of dit] (dec) {semantic-acoustic\\decoder\\{\scriptsize\color{gray}4096$\times$ upsampling}};

\node[wave, right=0.6cm of dec] (wav_out) {
\begin{tikzpicture}[miniwave]

\draw[blue!70!black] plot[domain=0:12.56,samples=160]
(\x,{0.6*sin(2*\x r)});

\draw[red!70!black] plot[domain=0:12.56,samples=160]
(\x,{0.6*sin(2*(\x-0.4) r)});

\end{tikzpicture}\\[2pt]
{\scriptsize\color{gray} stereo 44.1\,kHz}
};

\draw[arr] (wav_in) -- (enc);
\draw[arr] (enc) -- (dit);
\draw[arr] (dit) -- (dec);
\draw[arr] (dec) -- (wav_out);

\node[cond] (text) at ([yshift=0.5cm, xshift=-2.0cm]dit.north)
{"a jazz piano solo"};

\node[cond] (dur) at ([yshift=0.9cm]dit.north)
{duration in seconds};

\node[cond] (paint) at ([yshift=0.5cm, xshift=2.0cm]dit.north)
{editing controls};

\draw[arr] (text.south)  -- (text.south  |- dit.north);
\draw[arr] (dur.south)   -- (dur.south   |- dit.north);
\draw[arr] (paint.south) -- (paint.south |- dit.north);

\end{tikzpicture}

\caption{Stable Audio 3 text-to-audio models support variable-length generation and editing via inpainting. }
\label{fig:simplified_overview}
\end{figure}
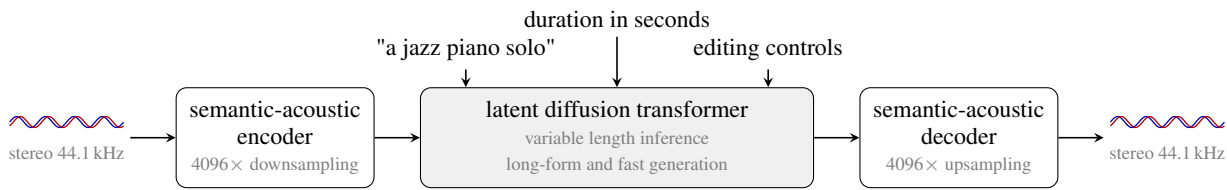

\begin{table}[h]
\centering
\begin{tabular}{lccccccc}
\toprule
 & maximum  & \# learnable & H200  & open  & licensed & \multicolumn{2}{c}{generation} \\
 & length & parameters & inference time  & weights & data & music & sfx \\
\midrule
\texttt{small-music}  & 2m        & 459M & 0.44s  & \checkmark & \checkmark & \checkmark & --- \\
\texttt{small-sfx}  & 2m        & 459M & 0.44s  & \checkmark & \checkmark & --- & \checkmark \\
\texttt{medium} & 6m 20s & 1.4B & {1.31s}  & \checkmark & \checkmark & \checkmark & \checkmark \\
\texttt{large}  & 6m 20s & 2.7B & 1.80s  & --- & \checkmark & \checkmark & \checkmark \\
\bottomrule
\end{tabular}
\vspace{3mm}
\caption{Stable Audio 3 models support the generation of long audio sequences while maintaining fast inference times. Parameter counts are for the diffusion transformer only. SAME-S and SAME-L have 108M and 852M parameters.}
\label{tab:model_family_intro}
\end{table}

\vspace{-3mm}

\section{Introduction}\label{sec:introduction}

Recent progress in music and audio generation has been driven by two broad families of models: autoregressive models~\cite{copet2023musicgen,agostinelli2023musiclm,yue2025,heartmula2026} and latent diffusion models~\cite{evans2024stableaudio,evans2024longform,liu2023audioldm,chen2024musicldm,schneider2024mousai,liu2024audioldm2}.
Autoregressive models have achieved strong results by operating sequentially on discrete audio tokens. In contrast, latent diffusion models generate continuous latent representations that are subsequently decoded with a separate autoencoder, offering an alternative that avoids discrete tokenisation and autoregressive sampling. Complementing these approaches, hybrid methods have been proposed using an autoregressive model to produce tokens that are then refined by a diffusion model~\cite{acestep2026,zhang2025inspiremusic}. Stable Audio 3 consists of three latent diffusion models at different scales (\texttt{small}, \texttt{medium}, \texttt{large}, see  Table~\ref{tab:model_family_intro}).

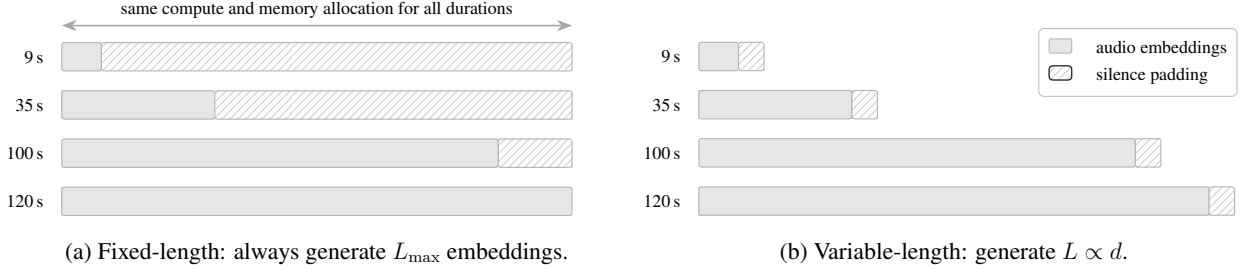
\begin{figure}[h]
\centering
\tikzset{
  >=Stealth,
}

\begin{minipage}[t]{0.48\textwidth}
\centering
\begin{tikzpicture}[scale=0.75, transform shape]

  \draw[<->, semithick, gray!70] (0,3.1) -- node[above, font=\small, black] {same compute and memory allocation for all durations} (9,3.1);

  \node[font=\footnotesize, anchor=east] at (-0.2,2.55) {9\,s};
  \fill[black!10] (0,2.3) rectangle ++(0.7,0.5);
  \draw[gray!60, rounded corners=1pt] (0,2.3) rectangle ++(0.7,0.5);
  \fill[white] (0.7,2.3) rectangle ++(8.3,0.5);
  \fill[pattern=north east lines, pattern color=gray!35] (0.7,2.3) rectangle ++(8.3,0.5);
  \draw[gray!60, rounded corners=1pt] (0.7,2.3) rectangle ++(8.3,0.5);

  \node[font=\footnotesize, anchor=east] at (-0.2,1.7) {35\,s};
  \fill[black!10] (0,1.45) rectangle ++(2.7,0.5);
  \draw[gray!60, rounded corners=1pt] (0,1.45) rectangle ++(2.7,0.5);
  \fill[white] (2.7,1.45) rectangle ++(6.3,0.5);
  \fill[pattern=north east lines, pattern color=gray!35] (2.7,1.45) rectangle ++(6.3,0.5);
  \draw[gray!60, rounded corners=1pt] (2.7,1.45) rectangle ++(6.3,0.5);

  \node[font=\footnotesize, anchor=east] at (-0.2,0.85) {100\,s};
  \fill[black!10] (0,0.6) rectangle ++(7.7,0.5);
  \draw[gray!60, rounded corners=1pt] (0,0.6) rectangle ++(7.7,0.5);
  \fill[white] (7.7,0.6) rectangle ++(1.3,0.5);
  \fill[pattern=north east lines, pattern color=gray!35] (7.7,0.6) rectangle ++(1.3,0.5);
  \draw[gray!60, rounded corners=1pt] (7.7,0.6) rectangle ++(1.3,0.5);

  \node[font=\footnotesize, anchor=east] at (-0.2,0.0) {120\,s};
  \fill[black!10] (0,-0.25) rectangle ++(9.0,0.5);
  \draw[gray!60, rounded corners=1pt] (0,-0.25) rectangle ++(9.0,0.5);

  \node[font=\large, anchor=north] at (4.5,-0.6) {(a) Fixed-length: always generate $L_{\max}$ embeddings.};

\end{tikzpicture}
\end{minipage}
\hfill
\begin{minipage}[t]{0.48\textwidth}
\centering
\begin{tikzpicture}[scale=0.75, transform shape]

  \node[draw=gray!50, rounded corners=2pt, fill=white, inner sep=5pt,
        font=\footnotesize, anchor=north east] at (9.5,3.05) {
    \begin{tabular}{@{}cl@{}}
      \tikz\draw[fill=black!10, draw=gray!60, rounded corners=1pt] (0,0) rectangle (0.4,0.25); &
        audio embeddings \\[3pt]
      \tikz{\draw[fill=white, draw=gray!60, rounded corners=1pt] (0,0) rectangle (0.4,0.25);
            \draw[pattern=north east lines, pattern color=gray!35] (0,0) rectangle (0.4,0.25);} &
        silence padding \\
    \end{tabular}
  };

  \node[font=\footnotesize, anchor=east] at (-0.2,2.55) {9\,s};
  \fill[black!10] (0,2.3) rectangle ++(0.7,0.5);
  \draw[gray!60, rounded corners=1pt] (0,2.3) rectangle ++(0.7,0.5);
  \fill[white] (0.7,2.3) rectangle ++(0.45,0.5);
  \fill[pattern=north east lines, pattern color=gray!35] (0.7,2.3) rectangle ++(0.45,0.5);
  \draw[gray!60, rounded corners=1pt] (0.7,2.3) rectangle ++(0.45,0.5);

  \node[font=\footnotesize, anchor=east] at (-0.2,1.7) {35\,s};
  \fill[black!10] (0,1.45) rectangle ++(2.7,0.5);
  \draw[gray!60, rounded corners=1pt] (0,1.45) rectangle ++(2.7,0.5);
  \fill[white] (2.7,1.45) rectangle ++(0.45,0.5);
  \fill[pattern=north east lines, pattern color=gray!35] (2.7,1.45) rectangle ++(0.45,0.5);
  \draw[gray!60, rounded corners=1pt] (2.7,1.45) rectangle ++(0.45,0.5);

  \node[font=\footnotesize, anchor=east] at (-0.2,0.85) {100\,s};
  \fill[black!10] (0,0.6) rectangle ++(7.7,0.5);
  \draw[gray!60, rounded corners=1pt] (0,0.6) rectangle ++(7.7,0.5);
  \fill[white] (7.7,0.6) rectangle ++(0.45,0.5);
  \fill[pattern=north east lines, pattern color=gray!35] (7.7,0.6) rectangle ++(0.45,0.5);
  \draw[gray!60, rounded corners=1pt] (7.7,0.6) rectangle ++(0.45,0.5);

  \node[font=\footnotesize, anchor=east] at (-0.2,0.0) {120\,s};
  \fill[black!10] (0,-0.25) rectangle ++(9.0,0.5);
  \draw[gray!60, rounded corners=1pt] (0,-0.25) rectangle ++(9.0,0.5);
  \fill[white] (9.0,-0.25) rectangle ++(0.45,0.5);
  \fill[pattern=north east lines, pattern color=gray!35] (9.0,-0.25) rectangle ++(0.45,0.5);
  \draw[gray!60, rounded corners=1pt] (9.0,-0.25) rectangle ++(0.45,0.5);

  \node[font=\large, anchor=north] at (4.5,-0.6) {(b) Variable-length: generate $L \propto d$.};

\end{tikzpicture}
\end{minipage}

\caption{Fixed- vs. variable-length generation. (a)~Fixed-length generation allocates $L_{\max}$ embeddings regardless of the requested duration $d$, wasting computation on zero-padded silence for short clips. (b)~Variable-length generation allocates $L$ embeddings that are proportional to the requested duration $d$ (silence padding is also used, see Section~\ref{subsec:varlen}).}
\label{fig:variable-length}
\end{figure}

Variable-length generation is a key capability of Stable Audio 3, particularly because our models generate very long audio (Table~\ref{tab:model_family_intro}). While autoregressive models naturally support variable-length outputs due to their sequential nature, diffusion models typically require generating the entire audio length at once~\cite{evans2024stableaudio,evans2024longform} \mbox{(Figure~\ref{fig:variable-length}: a).} This means that, \textit{e.g.}, generating a short sample with \texttt{small-music} would require producing a 2m audio with mostly silence. To address this compute and memory inefficiency, Stable Audio 3 supports variable-length generation \mbox{(Figure~\ref{fig:variable-length}: b)}, enabling efficient synthesis without incurring full-length computation for short outputs. Such efficiency gains are critical for deploying open-weight models on {consumer-grade hardware}, where compute and memory budgets are constrained.

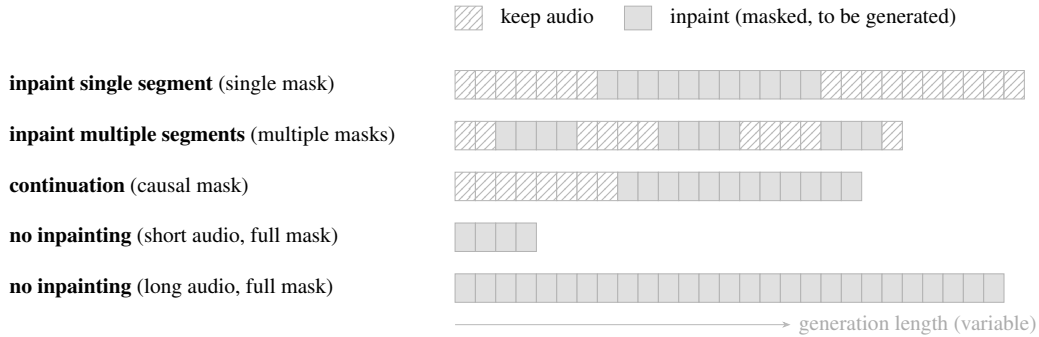
\begin{figure}[h]
\centering
\resizebox{0.85\columnwidth}{!}{
\usetikzlibrary{arrows.meta, patterns}

\begin{tikzpicture}[font=\small]

\def\cw{0.30}
\def\ch{0.42}
\def\rowgap{0.75}
\def\xlabel{-3.2}
\def\yshift{0.35}

\def\ylegend{0}
\def\yrone{-1*\rowgap + \yshift}
\def\yrtwo{-2*\rowgap + \yshift}
\def\yrfour{-3*\rowgap + \yshift}
\def\yrfive{-4*\rowgap + \yshift}
\def\yrsix{-5*\rowgap + \yshift}
\def\ytime{-6*\rowgap + \yshift}

\fill[white, draw=gray!60, line width=0.4pt] (0,\ylegend+0.2) rectangle (0.40,\ylegend+0.55);
\fill[pattern=north east lines, pattern color=gray!55] (0,\ylegend+0.2) rectangle (0.40,\ylegend+0.55);
\node[anchor=west] at (0.55,\ylegend+0.375) {keep audio};

\fill[black!12, draw=gray!60, line width=0.4pt] (2.5,\ylegend+0.2) rectangle (2.90,\ylegend+0.55);
\node[anchor=west] at (3.05,\ylegend+0.375) {inpaint (masked, to be generated)};

\node[anchor=west] at (\xlabel-3.5,\yrone-0.21)  {\textbf{inpaint single segment} (single mask)};
\node[anchor=west] at (\xlabel-3.5,\yrtwo-0.21)  {\textbf{inpaint multiple segments} (multiple masks)};
\node[anchor=west] at (\xlabel-3.5,\yrfour-0.21) {\textbf{continuation} (causal mask)};
\node[anchor=west] at (\xlabel-3.5,\yrfive-0.21) {\textbf{no inpainting} (short audio, full mask)};
\node[anchor=west] at (\xlabel-3.5,\yrsix-0.21)  {\textbf{no inpainting} (long audio, full mask)};

\foreach \i/\v in {
0/1,1/1,2/1,3/1,4/1,5/1,6/1,
7/0,8/0,9/0,10/0,11/0,12/0,13/0,14/0,15/0,16/0,17/0,
18/1,19/1,20/1,21/1,22/1,23/1,24/1,25/1,26/1,27/1}{
  \ifnum\v=1
    \fill[white, draw=gray!60, line width=0.4pt]
      (\i*\cw,\yrone) rectangle (\i*\cw+\cw,\yrone-\ch);
    \fill[pattern=north east lines, pattern color=gray!55]
      (\i*\cw,\yrone) rectangle (\i*\cw+\cw,\yrone-\ch);
  \else
    \fill[black!12, draw=gray!60, line width=0.4pt]
      (\i*\cw,\yrone) rectangle (\i*\cw+\cw,\yrone-\ch);
  \fi
}

\foreach \i/\v in {
0/1,1/1,
2/0,3/0,4/0,5/0,
6/1,7/1,8/1,9/1,
10/0,11/0,12/0,13/0,
14/1,15/1,16/1,17/1,
18/0,19/0,20/0,
21/1}{
  \ifnum\v=1
    \fill[white, draw=gray!60, line width=0.4pt]
      (\i*\cw,\yrtwo) rectangle (\i*\cw+\cw,\yrtwo-\ch);
    \fill[pattern=north east lines, pattern color=gray!55]
      (\i*\cw,\yrtwo) rectangle (\i*\cw+\cw,\yrtwo-\ch);
  \else
    \fill[black!12, draw=gray!60, line width=0.4pt]
      (\i*\cw,\yrtwo) rectangle (\i*\cw+\cw,\yrtwo-\ch);
  \fi}

\foreach \i in {0,...,7} {
  \fill[white, draw=gray!60, line width=0.4pt]
    (\i*\cw,\yrfour) rectangle (\i*\cw+\cw,\yrfour-\ch);
  \fill[pattern=north east lines, pattern color=gray!55]
    (\i*\cw,\yrfour) rectangle (\i*\cw+\cw,\yrfour-\ch);
}

\foreach \i in {8,...,19} {
  \fill[black!12, draw=gray!60, line width=0.4pt]
    (\i*\cw,\yrfour) rectangle (\i*\cw+\cw,\yrfour-\ch);
}

\foreach \i in {0,...,3} {
  \fill[black!12, draw=gray!60, line width=0.4pt]
    (\i*\cw,\yrfive) rectangle (\i*\cw+\cw,\yrfive-\ch);
}

\foreach \i in {0,...,26} {
  \fill[black!12, draw=gray!60, line width=0.4pt]
    (\i*\cw,\yrsix) rectangle (\i*\cw+\cw,\yrsix-\ch);
}

\draw[-{Stealth[length=3.5pt]}, gray!55, thin]
  (0,\ytime) -- (16.5*\cw,\ytime)
  node[anchor=west, font=\small, gray!65] {generation length (variable)};

\end{tikzpicture}
}
\caption{Editing with inpainting. Users provide audio and specify target segments for editing (gray, masked) while preserving original audio (hatched). This enables tasks from single or multi-segment editing to causal continuation.}
\label{fig:inpainting_masks}
\end{figure}

Controllability is also an important feature of modern generative audio and music models. Stable Audio 3 includes inpainting capabilities that allow editing targeted segments of audio, such as modifying a single segment (Figure \ref{fig:inpainting_masks}: first row), performing multi-segment edits (Figure \ref{fig:inpainting_masks}: second row), or supporting continuation (Figure \ref{fig:inpainting_masks}, third row), where the model can extend a given audio coherently beyond its original endpoint. This enables applications such as transient editing in percussive sounds, generating ideas for an unfinished song, or the extension of short recordings.

Stable Audio 3 comprises latent diffusion models built on top of a semantic-acoustic autoencoder. This latent representation is designed to preserve reconstruction fidelity while remaining generatively tractable and semantically structured for downstream use. Our aim is to maintain high-fidelity audio reconstruction while learning a compact latent space (with 4096$\times$ downsampling) that is both easy to model generatively with diffusion and structured in a semantically meaningful way. Specifically, acoustic fidelity is enforced using spectral reconstruction losses and adversarial training~\cite{defossez2023encodec,kumar2023dac}, while semantic structure is induced through latent-space regression objectives, including chroma and interaural level difference regression.
One important characterisic of the employed autoencoder is its $4096\times$ downsampling ratio, substantially higher than the $1024$ to $2048\times$ ratios common in prior work~\cite{defossez2023encodec,kumar2023dac,wang2025back}. This aggressive downsampling is central to our goals: it reduces sequence lengths enough for \texttt{medium} and \texttt{small} to generate long-form music and sound effects on consumer-grade GPUs and on a MacBook Pro using CPU.

Diffusion models typically require several inference steps to generate high-quality outputs, as they progressively refine noise through iterative denoising~\cite{ho2020ddpm,song2021sde}. Yet, fast inference is essential for responsive creative tools to feel engaging and inspiring. To address this, we use adversarial post-training, which allows reducing the number of sampling steps while maintaining (or improving) output quality \cite{novack2025arc}. Overall, our latent diffusion training pipeline consists of three stages: flow matching pre-training~\cite{liu2022flow,lipman2023flow,albergo2023stochastic}, ODE warmup distillation~\cite{salimans2022progressive,luhman2021distillation}, and adversarial post-training \cite{novack2025arc}.

\begin{figure}[h]
\centering
\resizebox{0.65\columnwidth}{!}{
\begin{tikzpicture}[
    >=stealth,
    block/.style={rectangle, draw, rounded corners=4pt, minimum height=0.55cm, inner sep=4pt, font=\footnotesize, align=center},
    wave/.style={font=\footnotesize, align=center, inner sep=2pt},
    arr/.style={->, semithick},
    dim/.style={font=\scriptsize, above=8pt, midway},
    miniwave/.style={
        baseline=-0.5ex,
        scale=0.12,
        line width=0.5pt
    }
]

\node[wave] (audio_in) {
    \begin{tikzpicture}[miniwave]
        \draw[blue!70!black] plot[domain=0:12.56,samples=160] (\x,{0.6*sin(2*\x r)});
        \draw[red!70!black] plot[domain=0:12.56,samples=160] (\x,{0.6*sin(2*(\x-0.4) r)});
    \end{tikzpicture}\\[1pt]
    {\scriptsize\color{gray} stereo 44.1\,kHz}
};

\node[block, right=0.25cm of audio_in] (patch_enc) {patch};
\node[block, right=0.4cm of patch_enc] (enc) {encoder\\[-1pt]TRB};

\node[block, right=0.4cm of enc] (bn) {soft-norm\\[-1pt]bottleneck};
\node[font=\scriptsize, color=gray, above=1pt of bn] (ds) {4096$\times$ downsampling};

\node[block, right=0.4cm of bn] (dec) {decoder\\[-1pt]TRB};
\node[block, right=0.4cm of dec] (patch_dec) {unpatch};

\node[wave, right=0.25cm of patch_dec] (audio_out) {
    \begin{tikzpicture}[miniwave]
        \draw[blue!70!black] plot[domain=0:12.56,samples=160] (\x,{0.6*sin(2*\x r)});
        \draw[red!70!black] plot[domain=0:12.56,samples=160] (\x,{0.6*sin(2*(\x-0.4) r)});
    \end{tikzpicture}\\[1pt]
    {\scriptsize\color{gray} stereo 44.1\,kHz}
};

\draw[arr] (audio_in) -- (patch_enc);
\draw[arr] (patch_enc) -- (enc);
\draw[arr] (enc) -- (bn);
\draw[arr] (bn) -- (dec);
\draw[arr] (dec) -- (patch_dec);
\draw[arr] (patch_dec) -- (audio_out);

\end{tikzpicture}
}
\caption{SAME autoencoder~\cite{parker2025same}. Stereo audio is reshaped into patch embededdings ($256\!\times$ downsampling), downsampled by an encoder TRB (further $16\!\times$), and passed through a soft-normalisation bottleneck with projection to latent dimension $d$. Latents are then reconstructed by a decoder TRB and unpatching. Total downsampling: $4096\!\times$.}
\label{fig:architecture}
\end{figure}
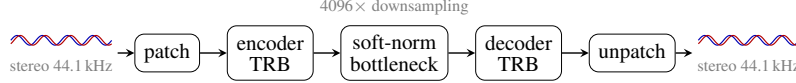

Stable Audio 3 is designed for broad community adoption as it is trained on licensed and Creative Commons data, enabling artists and developers to use it without legal concerns. Further, our models can scale from datacenter GPUs (\textit{e.g.}, H200) down to consumer-grade GPUs and even a MacBook Pro. The main contributions of Stable Audio 3 are:

\begin{itemize}

    \item Release the weights for \texttt{small} and \texttt{medium}, suitable to run on consumer-grade hardware (Table~\ref{tab:model_family_intro}).

    \item State-of-the-art results for text-to-audio generation for instrumental music and sounds (Section \ref{sec:discussion}).
    
    \item Fast inference: less than 2s to generate up to 6m 20s on an H200 (Sections \ref{sec:eval_music} and \ref{sec:eval_sfx}).

  \item Audio editing via inpainting, including single- and multi-segment edits and continuation (Section \ref{sec:eval_editing}).

    \item Propose a new method for variable-length audio generation with latent diffusion models (Section \ref{subsec:varlen}).

\item Several technical innovations:
a semantic-acoustic autoencoder that learns a compact latent for diffusion by preserving high-fidelity reconstruction and semantic information (Section~\ref{subsec:same}); the use of Transformer Resampling Blocks (TRBs, Section \ref{subsec:same}) for down/up-sampling;
a diffusion transformer improved with differential attention~\cite{ye2025diffattn}, adaptive layer normalization conditioning~\cite{peebles2023scalable}, and memory embeddings~\cite{burtsev2020memory, darcet2024registers}~(Section~\ref{fig:architecture});
minibatch optimal transport coupling for flow matching training~(Section~\ref{subsec:flow_matching});
and a distillation warmup stage followed by adversarial post-training for improved few-step generation~(Sections~\ref{subsec:ode_warmup} and~\ref{subsec:arc}).

\end{itemize}

\subsection{Related Work}\label{subsec:related}

\paragraph{Open models.} Early open models were predominantly based on either autoregressive approaches~\cite{copet2023musicgen,kreuk2023audiogen} or latent diffusion methods~\cite{liu2023audioldm,chen2024musicldm,liu2024audioldm2,ghosal2023tango,majumder2024tango2,evans2024stableaudioopen,ning2025diffrhythm}.
More recent open models continue to explore autoregressive approaches~\cite{yue2025,heartmula2026}, while also introducing flow matching methods~\cite{novack2025arc,jiang2026diffrhythm2,jam2025} and hybrid architectures that combine autoregressive modeling with flow matching~\cite{zhang2025inspiremusic,acestep2026,jiang2026diffrhythm2}.
These models have been applied to a range of audio generation tasks, including instrumental music~\cite{chen2024musicldm,copet2023musicgen, liu2024audioldm2}, sound effects~\cite{liu2023audioldm,liu2024audioldm2,novack2025arc,kreuk2023audiogen,ghosal2023tango,evans2024stableaudioopen}, and songs with vocals~\cite{yue2025,heartmula2026,acestep2026,zhang2025inspiremusic,ning2025diffrhythm,jam2025,jiang2026diffrhythm2}. Stable Audio 3 is a family of open-weight models (\texttt{small}, \texttt{medium}) based on flow matching for instrumental music and sound effects generation. For evaluation, we compare against the most competitive open models available.

\paragraph{Variable length.} Autoregressive models naturally support variable-length generation by producing tokens sequentially until an end-of-sequence token is produced, making variable length generation an emergent property. In contrast, latent diffusion models are typically defined over fixed-length sequences, requiring shorter inputs to be padded~\cite{evans2024longform,evans2024stableaudio}. This ties inference cost to a predefined maximum length rather than the actual content, leading to inefficiencies and limiting its practical scalability to long-form generation.
A similar issue has been addressed in image diffusion: early models~\cite{podell2023sdxl} relied on resolution conditioning and cropping to handle varying sizes, whereas modern transformer-based approaches rely on positional encodings to digest inputs of various sizes organically~\cite{chen2024pixart,li2024hunyuan}. Audio diffusion is beginning to follow this shift with approaches like autoregressive block-wise diffusion~\cite{jiang2026diffrhythm2}. Yet, fully native variable-length audio generation with diffusion remains largely unaddressed. To our knowledge, Stable Audio 3 models are the first to tackle this challenge in a manner analogous to recent advances in image diffusion.

\paragraph{Semantic latent spaces.} Most latent diffusion models operate on low-dimensional (64, 32) latents from VAEs trained focusing on acoustic reconstruction \cite{kreuk2023audiogen,evans2024stableaudioopen}.
Representation autoencoders (RAE)~\cite{zheng2025rae,li2026scalingrae} have shown that diffusion in higher-dimensional, semantically structured latent spaces yields faster convergence and better generation quality in the image domain. To our knowledge, Stable Audio~3 models are the first to explore this idea in the audio domain by relying on the Semantically-Aligned Music autoEncoder (SAME)~\cite{parker2025same}, which produces 256-dim latents designed to encode both acoustic fidelity and high-level semantic structure at a high downsampling ratio (4096$\times$).

\paragraph{Controllability.}
The demand for controllable audio generation is increasing as creative workflows require control beyond prompts~\cite{pons2025music}. Prior work can be categorized as follows: mask-based, instruction-based, inference-time control, global control, time-varying control, and lyrics editing.
Mask-based methods enable localized editing or continuation by generating the masked segments of a given audio~\cite{garcia2023vampnet,li2024jen1,seetharaman2026generative}. Instruction-based approaches support operations such as adding, removing, processing, or replacing sound sources through structured commands~\cite{wang2023audit,han2024instructme,parker2024stemgen}. Inference-time controls include guidance-based and inversion methods~\cite{novack2025arc,levy2023controllable,novack2026low,lan2024high,novack2024ditto,novack2024ditto2}. Global conditioning methods generate audio based on a reference signal~\cite{Tal2024JointAA,rouard2024audio} while time-varying controls introduce temporally dynamic constraints~\cite{Wu2023MusicCM, garcia2025sketch2sound,wang2025audio,kim2025enhancing}. Lyrics editing allows additional control by modifying textual content~\cite{heartmula2026,acestep2026,jam2025}.
Stable Audio 3 focuses on mask-based editing, as it does not require additional training data annotation. Training instead relies on simple random and causal masking. We do not consider instruction-based approaches, which typically require training datasets with stems, nor lyrics editing, which lies outside the scope of our work. We also exclude inference-time, global, and time-varying controls, as these often rely on model fine-tuning (LoRA~\cite{kim2025enhancing,wang2025audio}) or auxiliary models (ControlNet~\cite{Wu2023MusicCM}). Note that such controls can be included by fine-tuning Stable Audio 3 after its release.

\paragraph{Few-step generation.} The iterative denoising process of diffusion incurs high inference latency, motivating few-step generation methods. Reducing the number of sampling steps in diffusion can be achieved through distillation~\cite{salimans2022progressive,song2023consistency} or adversarial approaches~\cite{xiao2022ddgan,sauer2024add}.
In (step) distillation approaches the teacher provides direct supervision to train a distilled few-step generator that learns to map multiple inference steps into a single step, or a small number of steps, by distilling the teacher’s trajectories. However, most distillation approaches come with practical drawbacks like
online methods~\cite{Ren2024HyperSDTS, Wang2024PhasedCM, song2023consistency, lu2024simplifying, chen2025sana, Kim2023ConsistencyTM, Novack2025Presto, xu2025one, yin2024improved}, which are costly to train as they require 2-3 full models held in memory, or offline methods~\cite{liu2022flow, salimans2022progressive, kang2024distilling}, which require significant resources to generate and store trajectories to later train on.
To avoid such drawbacks, some explored adversarial post-training (without distillation)~\cite{xu2024ufogen, lin2025diffusion}. These works are {primarily} adversarial, as opposed to distillation methods that use adversarial auxiliary losses~\cite{yin2024improved,Novack2025Presto,Kim2023ConsistencyTM,Sauer2024FastHI,sauer2024add}, and use {real} data rather than teacher-generated samples, thus freeing the costly requirement of using trajectories. The adversarial loss encourages realism, making each estimate better than the standard distilled estimates. Such improved estimates enable post-trained models to use fewer sampling steps~\cite{xu2024ufogen, lin2025diffusion}.
In the audio domain: AudioLCM~\cite{liu2024audiolcm} used latent consistency distillation, Presto~\cite{Novack2025Presto} combined step and layer distillation, ARC~\cite{novack2025arc} combined relativistic and contrastive adversarial losses, and Woosh used MeanFlow distillation \cite{hadjeres2026woosh}.
Stable Audio 3 is based on ARC adversarial post-training but also uses distillation as a warmup.

\section{Architecture}\label{sec:architecture}

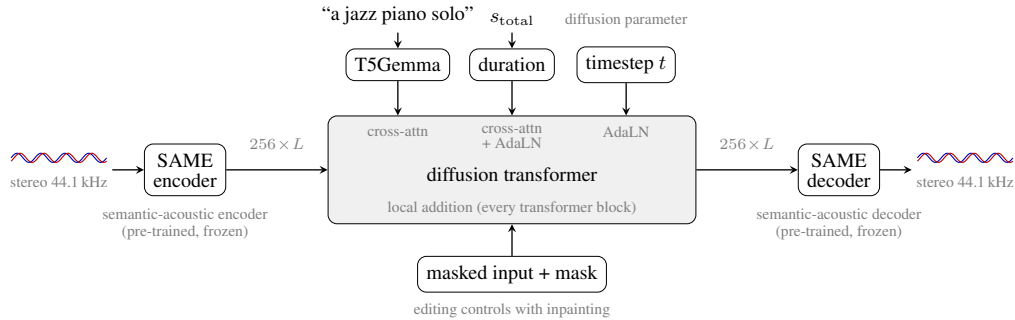
\begin{figure}[h]
\centering
\vspace{-1mm}
\resizebox{0.82\columnwidth}{!}{
\begin{tikzpicture}[
    >=stealth,
    block/.style={rectangle, draw, rounded corners=4pt, minimum height=0.55cm, inner sep=4pt, font=\footnotesize, align=center},
    wave/.style={font=\footnotesize, align=center, inner sep=2pt},
    arr/.style={->, semithick},
    dim/.style={font=\scriptsize, above=8pt, midway},
    cond/.style={rectangle, draw, rounded corners=4pt, minimum height=0.5cm, inner sep=4pt, font=\footnotesize, align=center},
    annot/.style={font=\scriptsize, color=gray},
]

\tikzset{miniwave/.style={
    baseline=-0.5ex,
    scale=0.12,
    line width=0.5pt
}}

\node[wave] (wav_in) {
\begin{tikzpicture}[miniwave]
\draw[blue!70!black] plot[domain=0:12.56,samples=160] (\x,{0.6*sin(2*\x r)});
\draw[red!70!black] plot[domain=0:12.56,samples=160] (\x,{0.6*sin(2*(\x-0.4) r)});
\end{tikzpicture}\\[1pt]
{\scriptsize\color{gray} stereo 44.1\,kHz}
};

\node[block, right=0.5cm of wav_in] (enc) {SAME\\[-1pt]encoder};

\node[block, fill=gray!12, right=1.6cm of enc, minimum width=5.8cm, minimum height=1.7cm] (dit) {\vspace{8.5cm}\\diffusion transformer};

\node[block, right=1.6cm of dit] (dec) {SAME\\[-1pt]decoder};

\node[wave, right=0.5cm of dec] (wav_out) {
\begin{tikzpicture}[miniwave]
\draw[blue!70!black] plot[domain=0:12.56,samples=160] (\x,{0.6*sin(2*\x r)});
\draw[red!70!black] plot[domain=0:12.56,samples=160] (\x,{0.6*sin(2*(\x-0.4) r)});
\end{tikzpicture}\\[1pt]
{\scriptsize\color{gray} stereo 44.1\,kHz}
};

\draw[arr] (wav_in) -- (enc);
\draw[arr] (enc) -- node[dim, inner sep=0.1pt] {\color{gray}$256\!\times\!L$} (dit);
\draw[arr] (dit) -- node[dim, inner sep=0.1pt] {\color{gray}$256\!\times\!L$} (dec);
\draw[arr] (dec) -- (wav_out);

\node[annot, below=2pt of enc] {\shortstack{semantic-acoustic encoder\\(pre-trained, frozen)}};
\node[annot, below=2pt of dec] {\shortstack{semantic-acoustic decoder\\(pre-trained, frozen)}};

\node[cond] (text) at ([yshift=0.8cm, xshift=-1.8cm]dit.north) {T5Gemma};
\node[cond] (dur) at ([yshift=0.8cm]dit.north) {duration};
\node[annot] at ([yshift=1.5cm, xshift=1.8cm]dit.north) {diffusion parameter};
\node[cond] (tstep) at ([yshift=0.8cm, xshift=1.8cm]dit.north) {timestep $t$};

\node[font=\footnotesize, above=0.25cm of text] (prompt) {``a jazz piano solo''};
\node[font=\footnotesize, above=0.25cm of dur] (secs) {$s_{\mathrm{total}}$};
\draw[arr] (prompt) -- (text);
\draw[arr] (secs) -- (dur);

\draw[arr] (text.south) -- ([xshift=-1.8cm]dit.north);
\draw[arr] (dur.south) -- (dit.north);
\draw[arr] (tstep.south) -- ([xshift=1.8cm]dit.north);

\node[annot] at ([xshift=-1.8cm, yshift=-0.25cm]dit.north) {cross-attn};
\node[annot] at ([yshift=-0.33cm]dit.north) {\shortstack{cross-attn\\[-1pt]+ AdaLN}};
\node[annot] at ([xshift=1.8cm, yshift=-0.25cm]dit.north) {AdaLN};

\node[cond, below=0.5cm of dit] (inpaint) {masked input + mask};
\draw[arr] (inpaint) -- (dit.south);
\node[annot] at ([yshift=0.25cm]dit.south) {local addition (every transformer block)};
\node[annot, below=1.5pt of inpaint] (inpaint_annot) {editing controls with inpainting};

\end{tikzpicture}

}
\caption{Stereo audio at 44.1\,kHz is encoded to a 256-dim latent sequence by a SAME autoencoder (4096$\times$ downsampling). A diffusion transformer generates latent sequences conditioned on text embeddings from T5Gemma (via cross-attention), a duration embedding (via both cross-attention and AdaLN), and a diffusion timestep $t$ (via AdaLN). Inpainting conditioning (masked input of 256$\times$L is concatenated with a binary mask, resulting in a size of $257 \times L$) is projected to the model dimension and added at each transformer block. SAME decoder reconstructs the latents.}
\label{fig:system_overview}
\end{figure}

Stable Audio 3 consists of two components: a semantic-acoustic autoencoder that maps waveforms to and from a continuous latent space; and a diffusion transformer generating latent sequences that is conditioned on text prompts, duration information, and inpainting masks. Figure~\ref{fig:system_overview} depicts the overall system. Training details are in Section \ref{sec:training} and further implementation details are available online via our code release.

\vspace{-2mm}
\begin{table}[h]
\centering
\begin{tabular}{lcccccccc}
\toprule
 &  &  &  & differential & SAME & maximum & \# learnable \\
 & $d$ & $D$ & $H$ & attention & autoencoder & length & parameters \\
\midrule
\texttt{small} & 1024 & 20 & 16 & \textemdash & SAME-S & 2m    & 459M \\
\texttt{medium}  & 1536 & 24 & 24 & \checkmark  & SAME-L & 6m 20s & 1.4B \\
\texttt{large}  & 2048 & 26 & 32 & \checkmark  & SAME-L & 6m 20s  & 2.7B \\
\bottomrule
\end{tabular}
\vspace{4mm}
\caption{Stable Audio 3 models where $d$, $D$, and $H$ denote transformer hyperparameters: latent dimensionality, number of transformer blocks, and number of attention heads, respectively. \texttt{small} uses no differential attention. Parameter counts are for the diffusion transformer only. SAME-S and SAME-L have 108M and 852M parameters, respectively.}
\label{tab:sa3models}
\vspace{-4mm}
\end{table}

\subsection{Semantic-Acoustic Autoencoder}\label{subsec:same}

\begin{figure}[h]
\vspace{-4mm}
\centering
\resizebox{0.47\columnwidth}{!}{
\begin{tikzpicture}[
    >=stealth,
    tok/.style={rectangle, draw=gray!60, minimum width=0.4cm, minimum height=0.4cm, inner sep=0pt, font=\tiny},
    inp/.style={tok, fill=black!12},
    qtok/.style={tok, fill=black!28},
    ext/.style={tok, fill=black!28, line width=0.6pt},
    stage/.style={rectangle, draw=gray!60, rounded corners=4pt, fill=black!6, inner sep=5pt, font=\scriptsize, align=center},
    arr/.style={->, semithick, black},
    lbl/.style={font=\scriptsize},
    brace/.style={decorate, decoration={brace, amplitude=3pt}},
    brace_m/.style={decorate, decoration={brace, amplitude=3pt, mirror}},
]

\def\sp{0.55}  
\def\sg{1.25}  
\node[inp] (x0) at (0,0) {$x_0$};
\node[inp] (x1) at (\sp,0) {$x_1$};
\node[qtok] (q0) at (2*\sp,0) {$q$};
\pgfmathsetmacro{\sB}{2+\sg}
\node[inp] (x2) at (\sB*\sp,0) {$x_2$};
\node[inp] (x3) at ({(\sB+1)*\sp},0) {$x_3$};
\node[qtok] (q1) at ({(\sB+2)*\sp},0) {$q$};
\pgfmathsetmacro{\sC}{\sB+2+\sg}
\node[inp] (x4) at (\sC*\sp,0) {$x_4$};
\node[inp] (x5) at ({(\sC+1)*\sp},0) {$x_5$};
\node[qtok] (q2) at ({(\sC+2)*\sp},0) {$q$};
\pgfmathsetmacro{\sD}{\sC+2+\sg}
\node[inp] (x6) at (\sD*\sp,0) {$x_6$};
\node[inp] (x7) at ({(\sD+1)*\sp},0) {$x_7$};
\node[qtok] (q3) at ({(\sD+2)*\sp},0) {$q$};

\draw[brace] ([yshift=2pt]x0.north west) -- ([yshift=2pt]q0.north east) node[midway, above=3pt, font=\tiny] {segment\,1};
\draw[brace] ([yshift=2pt]x2.north west) -- ([yshift=2pt]q1.north east) node[midway, above=3pt, font=\tiny] {segment\,2};
\draw[brace] ([yshift=2pt]x4.north west) -- ([yshift=2pt]q2.north east) node[midway, above=3pt, font=\tiny] {segment\,3};
\draw[brace] ([yshift=2pt]x6.north west) -- ([yshift=2pt]q3.north east) node[midway, above=3pt, font=\tiny] {segment\,4};

\pgfmathsetmacro{\cen}{(\sD+2)*\sp/2}

\draw[arr] (\cen, -0.35) -- (\cen, -0.78);

\pgfmathsetmacro{\trbw}{(\sD+2)*\sp}
\pgfmathsetmacro{\tsp}{\trbw / 6}  
\node[rectangle, draw=gray!50, dashed, rounded corners=4pt, fill=black!3,
      minimum width=\trbw cm, minimum height=0.8cm] (trb) at (\cen, -1.25) {};
\node[lbl, font=\tiny, anchor=south west] at (trb.north west) {stack of transformer layers};
\node[stage, minimum width=0.7cm, minimum height=0.45cm] (t1) at (\cen - 2.5*\tsp, -1.25) {$\mathcal{T}_1$};
\node[stage, minimum width=0.7cm, minimum height=0.45cm] (t2) at (\cen - 1.5*\tsp, -1.25) {$\mathcal{T}_2$};
\node[stage, minimum width=0.7cm, minimum height=0.45cm] (t3) at (\cen - 0.5*\tsp, -1.25) {$\mathcal{T}_3$};
\node[stage, minimum width=0.7cm, minimum height=0.45cm] (t4) at (\cen + 0.5*\tsp, -1.25) {$\mathcal{T}_4$};
\node[font=\scriptsize] (tdots) at (\cen + 1.5*\tsp, -1.25) {$\cdots$};
\node[stage, minimum width=0.7cm, minimum height=0.45cm] (tD) at (\cen + 2.5*\tsp, -1.25) {$\mathcal{T}_D$};
\draw[arr] (t1) -- (t2);
\draw[arr] (t2) -- (t3);
\draw[arr] (t3) -- (t4);
\draw[arr] (t4) -- (tdots);
\draw[arr] (tdots) -- (tD);

\draw[arr] (\cen, -1.75) -- (\cen, -2.15);

\node[tok, draw=gray!45, text=gray!55, densely dotted] at (0,-2.5) {$x_0$};
\node[tok, draw=gray!45, text=gray!55, densely dotted] at (\sp,-2.5) {$x_1$};
\node[ext] (y0) at (2*\sp,-2.5) {$y_0$};
\node[tok, draw=gray!45, text=gray!55, densely dotted] at (\sB*\sp,-2.5) {$x_2$};
\node[tok, draw=gray!45, text=gray!55, densely dotted] at ({(\sB+1)*\sp},-2.5) {$x_3$};
\node[ext] (y1) at ({(\sB+2)*\sp},-2.5) {$y_1$};
\node[tok, draw=gray!45, text=gray!55, densely dotted] at (\sC*\sp,-2.5) {$x_4$};
\node[tok, draw=gray!45, text=gray!55, densely dotted] at ({(\sC+1)*\sp},-2.5) {$x_5$};
\node[ext] (y2) at ({(\sC+2)*\sp},-2.5) {$y_2$};
\node[tok, draw=gray!45, text=gray!55, densely dotted] at (\sD*\sp,-2.5) {$x_6$};
\node[tok, draw=gray!45, text=gray!55, densely dotted] at ({(\sD+1)*\sp},-2.5) {$x_7$};
\node[ext] (y3) at ({(\sD+2)*\sp},-2.5) {$y_3$};

\node[font=\tiny] at (\cen, -3.05) {keep $y$ embeddings (discard $x$)};

\end{tikzpicture}
}
\caption{Example of TRB using embedding interleaving for 2$\times$ downsampling. Inputs are segmented into groups of~2 and a learnable output embedding is appended to each segment. The full interleaved sequence is processed by $D$ transformer layers $\mathcal{T}_D$. Output embeddings $y$ are extracted (discard $x$) to form the downsampled representation.}
\label{fig:trb}
\vspace{-4mm}
\end{figure}
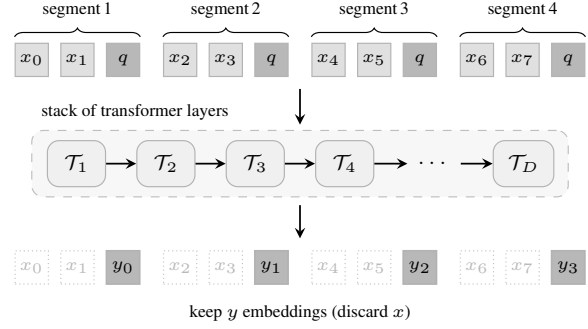

Our autoencoder builds on SAME~\cite{parker2025same}, a transformer-based autoencoder for audio that combines an initial patching stage with a Transformer Resampling Block (TRB, Figure \ref{fig:trb}).
Patching reshapes stereo audio into non-overlapping patches of 256 samples (per channel, resulting in $256\times$ downsampling).
TRB layers perform an additional $16\times$ downsampling by interleaving learnable output embeddings with the input sequence, and processing the resulting sequence with a stack of transformer layers using differential attention~\cite{ye2025diffattn} and rotary position embeddings~\cite{su2024roformer}.

The combined compression ratio is $4096\times$, yielding 256-dimensional latent embeddings at approximately 10.76\,Hz for 44.1\,kHz stereo input. Between encoder and decoder, a soft-normalisation bottleneck constrains the scale of the latent by using a learnable affine transform with running standard deviation tracking, providing a deterministic encoding.

For upsampling, the TRB process is reversed: each input embedding is paired with a number of output embeddings that are then extracted after processing. For example, to upsample 2$\times$, we interleave two output embeddings with each input embedding to retain the output embeddings after processing and discard the original input embeddings.

The SAME autoencoder is trained with a combination of (i) spectral reconstruction, (ii) adversarial, (iii) diffusion alignment, (iv) semantic regression, and (v) contrastive latent alignment losses that are designed to preserve reconstruction fidelity while remaining generatively tractable and semantically structured for downstream use. More specifically, SAME uses a multi-resolution STFT loss computed at seven resolutions (FFT sizes from 32 to 2048, each with 75\% overlap). A K-weighting pre-emphasis filter is applied before the STFT. At each resolution, the loss combines a spectral contrast term, a modified log-magnitude L1 distance, and an instantaneous frequency + group delay
(IFGD) phase loss \cite{parker2025same}. To handle stereo audio, the STFT loss is computed independently on both the sum-and-difference (mid/side) and per-channel (left/right) rep-resentations. Furthermore, the adversarial loss is formulated using a relativistic GAN objective. Then, the diffusion alignment loss consists of a small diffusion transformer (4 layers, 768-dimensional embeddings) that is trained jointly on the autoencoder’s latent space using a flow matching objective such that gradients flow back through the encoder, encouraging the latent geometry to be amenable to diffusion-based generation. SAME semantic regression losses include two lightweight linear regressors (single 1$\times$1 convolutions) to predict chroma and interaural level difference (ILD) features. Finally, the contrastive latent alignment loss employs a transformer-based critic (4 layers, 1024-dimensional) that is
trained to distinguish whether the latent sequence, wavelet (audio) features, and a T5Gemma text embedding (triplet) originate from the same input, encouraging the latent to preserve audio-level and cross-modal semantics. As a result, these losses focus on both high-quality acoustic reconstruction (spectral and adversarial losses) and semantic structure (semantic regression and contrastive latent alignment losses) for downstream diffusion (diffusion alignment loss).

The SAME autoencoder is frozen during diffusion training. \texttt{small} uses SAME-S, a distilled variant with fewer parameters (108M) designed for CPU inference, while \texttt{medium} and \texttt{large} use SAME-L (852M parameters).
Both variants share the same compression ratio and latent dimensionality. Further details are in the original SAME publication \cite{parker2025same}.

\subsection{Diffusion Transformer}

Our generative model is a diffusion transformer operating on SAME latents~\cite{peebles2023scalable}. 
Transformers replaced U-Nets for latent diffusion \cite{chen2024pixart}, and Stable Audio 3 adapts the diffusion transformer for text-to-audio with modifications including editing capabilities with inpainting, differential attention~\cite{ye2025diffattn}, memory embeddings \cite{burtsev2020memory}, and variable-length support.

\begin{figure}[h]
\centering
\vspace{-2mm}
\resizebox{0.9\textwidth}{!}{%
\begin{tikzpicture}[
    node distance=1.2cm,
    block/.style={rectangle, draw, rounded corners=4pt, minimum height=0.7cm, inner sep=4pt, font=\footnotesize, align=center},
    mem/.style={rectangle, draw, rounded corners=3pt, minimum height=0.55cm, inner sep=3pt, font=\scriptsize, align=center},
    op/.style={draw, circle, thick, minimum size=0.5cm, inner sep=0pt, font=\footnotesize},
    lbl/.style={font=\scriptsize, text=gray, align=center},
    dim/.style={font=\scriptsize, color=gray, midway, above=2pt},
    arr/.style={->, semithick, >=stealth},
]

\node[block] (in) {SAME latent\\[-1pt]{\scriptsize\color{gray}$256 {\times} T$}};

\node[block, right=0.4cm of in] (proj1) {linear\\[-1pt]{\scriptsize\color{gray}$256 {\to} d$}};

\node[op, right=1.4cm of proj1] (prep) {$\Vert$};
\node[mem, above=0.6cm of prep] (memin) {64 memory\\embeddings\\[2pt]{\scriptsize\color{gray}$d {\times} 64$}};

\node[block, right=1.5cm of prep] (tb) {transformer blocks $\times D$};

\node[op, right=1.5cm of tb] (rem) {$\setminus$};
\node[mem, above=0.6cm of rem] (memout) {discard\\memory\\embeddings};

\node[block, right=1.4cm of rem] (proj2) {linear\\[-1pt]{\scriptsize\color{gray}$d {\to} 256$}};

\node[block, right=0.4cm of proj2] (out) {SAME latent\\[-1pt]{\scriptsize\color{gray}$256 {\times} T$}};

\draw[arr] (in) -- (proj1);
\draw[arr] (proj1) -- node[dim] {$d {\times} T$} (prep);
\draw[arr] (memin) -- (prep);
\draw[arr] (prep) -- node[dim] {$d {\times} (T{+}64)$} (tb);
\draw[arr] (tb) -- node[dim] {$d {\times} (T{+}64)$} (rem);
\draw[arr] (rem) -- (memout);
\draw[arr] (rem) -- node[dim] {$d {\times} T$} (proj2);
\draw[arr] (proj2) -- (out);

\node[lbl, below=0.05cm of prep] {prepend};
\node[lbl, below=0.05cm of rem] {remove};

\end{tikzpicture}
}
\caption{Diffusion transformer architecture. SAME latents are linearly projected from $256$ to $d$ channels. A set of $64$ memory embeddings is prepended, providing a global memory buffer that every position can attend to. The resulting sequence is processed by $D$ transformer blocks with latent dimensionality $d$. After the final block, memory embeddings are discarded and the sequence is projected back to $256$ channels. In/out 1$\times$1 convolutions are omitted.}
\label{fig:dit_architecture}
\vspace{-4mm}
\end{figure}
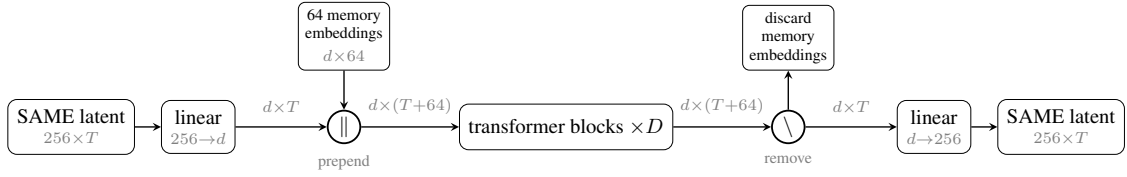

SAME latents first pass through a 1$\times$1 convolution with a residual connection. A linear projection then maps SAME frames to the transformer's latent dimensionality ($256 \to d$). Before entering the transformer, 64 learned memory embeddings are prepended. These embeddings serve as context that every position can attend to, effectively providing a global memory buffer. The resulting sequence is processed by a stack of $D$ transformer blocks with latent dimensionality $d$ and $H$ heads. After the final block, the memory embeddings are removed, and the sequence is projected back to the 256 dimensions of SAME. A final 1$\times$1 convolution with a residual connection produces the final output.

\begin{figure}[h]
\vspace{-6mm}
\centering
\begin{tikzpicture}[
  scale=0.7, transform shape,
  >=stealth,
  block/.style={draw=gray!60, rounded corners=3pt, minimum height=0.7cm,
                align=center, font=\small},
  selfattn/.style={block, fill=black!10},
  crossattn/.style={block, fill=black!6},
  ffn/.style={block, fill=black!16},
  norm/.style={block, fill=black!3},
  cond/.style={block, fill=black!20},
  gate/.style={block, fill=white},
  addcirc/.style={circle, draw=gray!60, inner sep=0pt, minimum size=5.5mm,
                  font=\scriptsize, fill=white},
  sdot/.style={circle, fill=black, inner sep=0pt, minimum size=2.5pt},
  dlabel/.style={font=\small, text=black, anchor=west, align=left},
  lin/.style={thick, black},
]

\begin{scope}[yshift=2.0cm]
  \def\R{-2.3}

  \node (Lin) at (0,0) {$\mathbf{x}$};
  \node[selfattn, minimum width=4.2cm] (Lsa) at (0,1.3) {self-attention};
  \node[addcirc] (La1) at (0,2.3) {$+$};
  \node[crossattn, minimum width=4.2cm] (Lca) at (0,3.4) {cross-attention};
  \node[addcirc] (La2) at (0,4.4) {$+$};
  \node[cond, minimum width=4.2cm] (Llc) at (0,5.5) {local-additive conditioning};
  \node[ffn,  minimum width=4.2cm] (Lff) at (0,6.8) {feed-forward network};
  \node[addcirc] (La3) at (0,7.8) {$+$};
  \node (Lout) at (0,8.6) {$\mathbf{x}'$};

  \coordinate (Ls1) at (0,0.6);
  \coordinate (Ls2) at (0,2.8);
  \coordinate (Ls3) at (0,6.2);

  \draw[lin] (Lin) -- (Lsa);
  \draw[lin] (Lsa) -- (La1);
  \draw[lin] (La1) -- (Lca);
  \draw[lin] (Lca) -- (La2);
  \draw[lin] (La2) -- (Llc);
  \draw[lin] (Llc) -- (Lff);
  \draw[lin] (Lff) -- (La3);
  \draw[lin] (La3) -- (Lout);

  \foreach \s/\a in {Ls1/La1, Ls2/La2, Ls3/La3}{
      \node[sdot] at (\s) {};
      \draw[lin, rounded corners=6pt] (\s) -- ++(\R,0) |- (\a.west);
  }

  \draw[dashed, rounded corners=4pt, gray!50]
      (\R-0.4,0.3) rectangle (2.3,8.3);
  \node[font=\scriptsize, gray!60, anchor=north west]
      at (\R-0.35,0.35) {$\times\,D$};
\end{scope}

\begin{scope}[xshift=8.5cm]
  \def\roff{-3.2}
  \def\Rw{5.0cm}
  \def\lx{3.6}

  \node (Rin) at (0,0) {$\mathbf{x}$};

  \node[norm, minimum width=\Rw] (Rn1) at (0,0.85)
      {RMSNorm};
  \node[dlabel] at (\lx,0.85) {RMS pre-norm};

  \node[gate, minimum width=\Rw] (Ra1) at (0,1.7)
      {$\mathbf{x}\!\cdot\!(1{+}\boldsymbol{\gamma}_s)
        +\boldsymbol{\beta}_s$};
  \node[dlabel] at (\lx,1.7) {AdaLN scale and shift};

  \node[selfattn, minimum width=\Rw] (Rsa) at (0,2.55)
      {self-attention};
  \node[dlabel] at (\lx,2.55) {QK-RMSNorm and partial RoPE};

  \node[gate, minimum width=\Rw] (Rg1) at (0,3.4)
      {$\mathbf{x}\!\cdot\!\sigma(1-\mathbf{g}_s)$};
  \node[dlabel] at (\lx,3.4) {AdaLN gate};

  \node[addcirc] (Rr1) at (0,4.1) {$+$};

  \draw[lin] (Rin)  -- (Rn1);
  \draw[lin] (Rn1)  -- (Ra1);
  \draw[lin] (Ra1)  -- (Rsa);
  \draw[lin] (Rsa)  -- (Rg1);
  \draw[lin] (Rg1)  -- (Rr1);
  \coordinate (Rs1) at (0,0.4);
  \node[sdot] at (Rs1) {};
  \draw[lin, rounded corners=6pt] (Rs1) -- ++(\roff,0) |- (Rr1.west);

  \node[norm, minimum width=\Rw] (Rn2) at (0,4.9)
      {RMSNorm};
  \node[dlabel] at (\lx,4.9) {RMS pre-norm};

  \node[crossattn, minimum width=\Rw] (Rca) at (0,5.75)
      {cross-attention};
  \node[dlabel] at (\lx,5.75) {QK-RMSNorm};

  \node[addcirc] (Rr2) at (0,6.45) {$+$};

  \draw[lin] (Rr1) -- (Rn2);
  \draw[lin] (Rn2) -- (Rca);
  \draw[lin] (Rca) -- (Rr2);
  \coordinate (Rs2) at (0,4.45);
  \node[sdot] at (Rs2) {};
  \draw[lin, rounded corners=6pt] (Rs2) -- ++(\roff,0) |- (Rr2.west);

  \node[cond, minimum width=\Rw] (Rlc) at (0,7.3)
      {local-additive conditioning};

  \draw[lin] (Rr2) -- (Rlc);

  \node[norm, minimum width=\Rw] (Rn3) at (0,8.55)
      {RMSNorm};
  \node[dlabel] at (\lx,8.55) {RMS pre-norm};

  \node[gate, minimum width=\Rw] (Ra2) at (0,9.4)
      {$\mathbf{x}\!\cdot\!(1{+}\boldsymbol{\gamma}_f)
        +\boldsymbol{\beta}_f$};
  \node[dlabel] at (\lx,9.4) {AdaLN scale and shift};

  \node[ffn,  minimum width=\Rw] (Rff) at (0,10.25)
      {SwiGLU FFN};
  \node[dlabel] at (\lx,10.25) {GLU with SiLU gate};

  \node[gate, minimum width=\Rw] (Rg2) at (0,11.1)
      {$\mathbf{x}\!\cdot\!\sigma(1-\mathbf{g}_f)$};
  \node[dlabel] at (\lx,11.1) {AdaLN gate};

  \node[addcirc] (Rr3) at (0,11.8) {$+$};

  \draw[lin] (Rlc) -- (Rn3);
  \draw[lin] (Rn3) -- (Ra2);
  \draw[lin] (Ra2) -- (Rff);
  \draw[lin] (Rff) -- (Rg2);
  \draw[lin] (Rg2) -- (Rr3);
  \coordinate (Rs3) at (0,8.1);
  \node[sdot] at (Rs3) {};
  \draw[lin, rounded corners=6pt] (Rs3) -- ++(\roff,0) |- (Rr3.west);

  \node (Rout) at (0,12.45) {$\mathbf{x}'$};
  \draw[lin] (Rr3) -- (Rout);
\end{scope}

\end{tikzpicture}
\vspace{-2mm}
\caption{High-level (left) and detailed (right) overview of a single transformer block. $\sigma$~denotes the sigmoid function. Adaptive layer normalization (AdaLN with gate, scale, and shift) is used for diffusion timestep and duration conditioning, cross-attention for text and duration conditioning, and local-additive conditioning for inpainting.}
\label{fig:dit-block}
\end{figure}
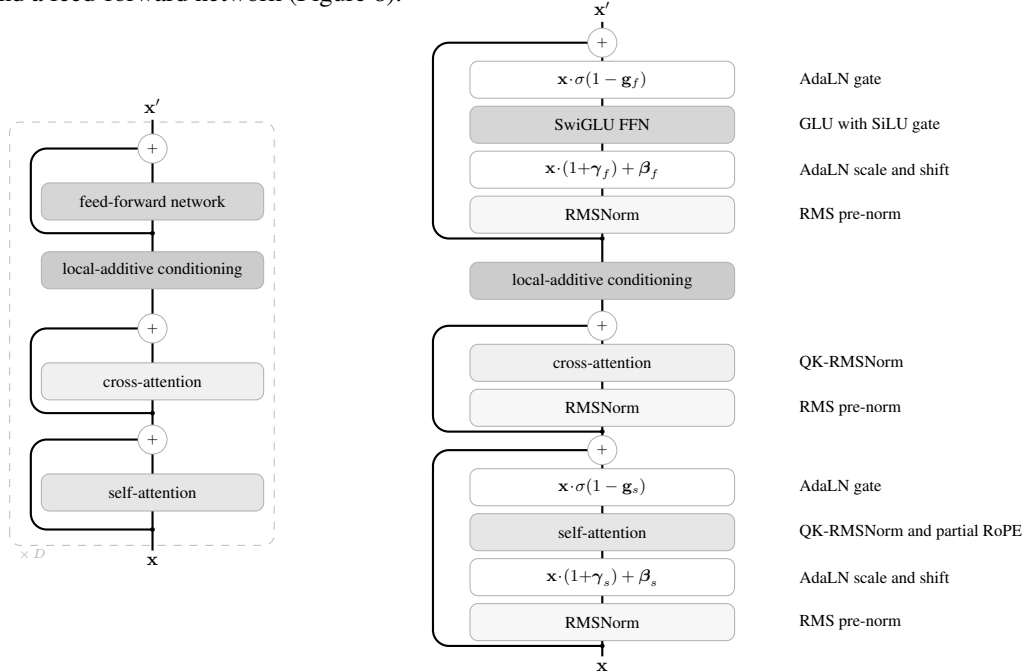

Conditioning information enters the transformer through three distinct pathways (Figure \ref{fig:dit-block}). First, the diffusion timestep and duration (length of generation) are mapped to a global embedding that modulates each self-attention and feed-forward layers in each transformer block with adaptive layer normalization (AdaLN). Second, text embeddings from a frozen T5Gemma encoder, concatenated with a duration embedding, employ cross-attention for conditioning. Third, for inpainting, a local-additive conditioning signal with the reference audio (to inpaint) and a binary mask (signaling where to inpaint) is projected through an MLP and added to the hidden state of each transformer block.

We train 3 models that share the same design but differ in transformer capacity, maximum generation length, and autoencoder (Table~\ref{tab:sa3models}). \texttt{medium} and \texttt{large} use differential attention~\cite{ye2025diffattn} in both self-attention and cross-attention layers, which roughly doubles the Q and K projection sizes relative to standard multi-head attention that \texttt{small} uses.

\paragraph{Transformer blocks.}
Each transformer block is composed of self-attention, cross-attention, local-additive conditioning, and a feed-forward network (Figure~\ref{fig:dit-block}).

Self-attention follows a pre-norm design with AdaLN \cite{peebles2023scalable,chen2024pixart,hadjeres2026woosh}. The input is normalised via RMSNorm \cite{liu2024deepseek}, then diffusion timestep and duration conditioning signals are injected (jointly) via AdaLN. In the following self-attention layer each head employs QK-RMSNorm to prevent dot-product outputs from growing unconstrained~\cite{henry2020query}. Positional embeddings are RoPE~\cite{su2024roformer} with partial rotation: only 32 of each head's dimensions are rotated, while the remainder carry no positional information. Finally, an AdaLN gate further conditions the output before the residual connection.

Cross-attention follows the same pre-norm design but without AdaLN. Embeddings are first normalised via RMSNorm and projected into queries, while keys and values are derived from the conditioning context (text and duration embeddings) through a separate projection. As in self-attention, each head also employs QK-RMSNorm~\cite{liu2024deepseek,henry2020query}. No positional embeddings are applied. The output embedding is added to the residual stream.

Local-additive conditioning enables inpainting by adding a frame-aligned signal before the feed-forward network. The inpainting conditioning signal (a binary mask concatenated with the masked reference audio) is projected through a 2-layer MLP with SiLU and added directly to the cross-attention output. MLP layers are zero-initialised, such that the inpainting pathway can be introduced into pretrained models without disrupting its learned representations.

The feed-forward network is a SwiGLU~\cite{shazeer2020glu} where the gated linear unit operates at 4$\times$ the model dimension $d$ and the gate is a swish (SiLU) gate instead of a sigmoid. After gating, a linear layer projects back from $4d$ to $d$. This part also uses RMS-based pre-norm and AdaLN as self-attention for diffusion timestep and duration conditioning.

\paragraph{Adaptive layer normalisation (AdaLN): diffusion timestep and duration conditioning.} 
The diffusion timestep $t \in [0,1]$ is mapped to a 256-dim Fourier features vector and then projected to $d$ by an MLP with SiLU.
The duration (in seconds) is normalised to $[0,1]$ and also encoded into a 256-dim Fourier features vector and then projected to $d$ by an MLP with SiLU.
These two $d$-dimensional embeddings are summed element-wise and passed through another MLP with SiLU that computes a shared conditioning embeddings that are fed to each transformer block.
As a result, every AdaLN ($\gamma_s$, $\beta_s$, $g_s$, $\gamma_f$, $\beta_f$, $g_f$) gets conditioning embeddings ($c_{\gamma,s}$, $c_{\beta,s}$, $c_{g,s}$, $c_{\gamma,f}$, $c_{\beta,f}$, $c_{g,f}$) that are shared across transformer blocks.
Finally, each transformer block independently learns 6 bias terms ($b_{\gamma,s}$, $b_{\beta,s}$, $b_{g,s}$, $b_{\gamma,f}$, $b_{\beta,f}$, $b_{g,f}$) that are added to the shared conditional embeddings to obtain the final AdaLN conditioning \cite{peebles2023scalable}. For example, the self-attention AdaLN scale is $\gamma_s = c_{\gamma,s} + b_{\gamma,s}$ where $c_{\gamma,s}$ is shared across blocks and $b_{\gamma,s}$ is block specific. The resulting AdaLN parameters at each transformer block ($\gamma_s$, $\beta_s$, $g_s$, $\gamma_f$, $\beta_f$, $g_f$) are applied following equations in Figure \ref{fig:dit-block}. This variant is referred to as {AdaLN-Single} \cite{chen2024pixart}, since the conditioning embeddings are shared across all transformer blocks, substantially reducing the number of conditioning parameters compared to standard AdaLN. Further, the multiplicative gating terms ($g_s$, $g_f$) are inspired by the modulation mechanism introduced in FLUX \cite{flux2024}.

\paragraph{Cross-attention: text and duration conditioning.}
Text is encoded by a T5Gemma (\texttt{google/t5gemma-b-b-ul2}) frozen encoder into a sequence of 256 embeddings of dimension 768. Short prompts are padded to 256 with a learned padding embedding, and long prompts are truncated to 256. The duration (in seconds) is normalised to $[0,1]$ and encoded into a 256-dim Fourier features vector and then projected to $d$ by an MLP with SiLU. These two conditioning sources are concatenated along the sequence dimension, forming a
context sequence of 257 embeddings. Text and duration conditioning enter each transformer block via cross-attention. Per-head QK-RMSNorm is applied to stabilize attention logits. Note
that duration conditioning thus enters each transformer block through two complementary pathways:
AdaLN together with the diffusion timestep, and cross-attention alongside the text prompt.

\paragraph{Local-additive conditioning for inpainting.} 
The region to edit with inpainting is signaled with a binary mask where ones mark frames to preserve and zeros mark frames to generate.
To that end, the original audio is encoded into latent space by the SAME autoencoder and element-wise multiplied by the mask, zeroing out the inpaint region. The single-channel mask and the 256-channel masked latent are concatenated along the channel dimension into a 257-dimensional per-frame conditioning tensor. In each transformer block, this tensor is projected to the transformer block dimension~$d$ by a MLP with SiLU and added element-wise to the residual stream between the cross-attention and feed-forward network. Because the output layer of the MLP is zero-initialized, the local-additive conditioning used for inpainting has no effect at the start of training, allowing smooth fine-tuning from a non-inpainting checkpoint.

\begin{figure}[h]
\centering
\resizebox{0.8\textwidth}{!}{%
\begin{tikzpicture}[
    node distance=0.2cm,
    block/.style={rectangle, draw, rounded corners=4pt, minimum height=0.55cm, inner sep=3pt, font=\footnotesize, align=center},
    cond/.style={rectangle, draw, rounded corners=4pt, minimum height=0.55cm, inner sep=3pt, font=\footnotesize, align=center},
    op/.style={draw, circle, thick, minimum size=0.45cm, inner sep=0pt, font=\footnotesize},
    lbl/.style={font=\scriptsize, text=gray, align=center},
    wave/.style={font=\footnotesize, align=center, inner sep=2pt},
    miniwave/.style={baseline=-0.5ex, scale=0.12, line width=0.5pt},
]

\node[wave] (wave) {
\begin{tikzpicture}[miniwave]
\draw[blue!70!black] plot[domain=0:12.56,samples=160] (\x,{0.6*sin(2*\x r)});
\draw[red!70!black] plot[domain=0:12.56,samples=160] (\x,{0.6*sin(2*(\x-0.4) r)});
\end{tikzpicture}\\[1pt]
{\scriptsize\color{gray} stereo 44.1\,kHz}
};

\node[block, right=of wave] (ae) {SAME\\[-1pt]{\scriptsize\color{gray} frozen}};
\node[block, right=of ae] (lat) {latent\\[-1pt]{\scriptsize\color{gray} $256 {\times} T$}};
\node[op, right=of lat] (mul) {$\odot$};
\node[cond, right=of mul] (mlat) {masked\\[-1pt]{\scriptsize\color{gray} $256 {\times} T$}};
\node[cond, right=of mlat] (cat) {concat\\[-1pt]{\scriptsize\color{gray} $257 {\times} T$}};
\node[cond, right=of cat] (mlp) {MLP w/ SiLU\\[-1pt]{\scriptsize\color{gray} $257 {\to} d{:}\; d{\times}T$}};
\node[block, right=of mlp] (pad) {left-pad\\[-1pt]{\scriptsize\color{gray} $257 {\times} (T{+}64)$}};
\node[op, right=of pad] (add) {$+$};

\node[block, above=0.7cm of mul] (mask) {mask\\[-1pt]{\scriptsize\color{gray} $1 {\times} T$}};

\node[lbl, above=0.9cm of add] (xemb) {cross-attention\\embedding};
\draw[->, semithick, >=stealth] (xemb) -- (add);

\node[lbl, right=0.3cm of add] (out) {feed-forward\\network};
\draw[->, semithick, >=stealth] (add) -- (out);

\draw[->, semithick, >=stealth] (wave) -- (ae);
\draw[->, semithick, >=stealth] (ae) -- (lat);
\draw[->, semithick, >=stealth] (lat) -- (mul);
\draw[->, semithick, >=stealth] (mul) -- (mlat);
\draw[->, semithick, >=stealth] (mlat) -- (cat);
\draw[->, semithick, >=stealth] (cat) -- (mlp);
\draw[->, semithick, >=stealth] (mlp) -- (pad);
\draw[->, semithick, >=stealth] (pad) -- (add);
\draw[->, semithick, >=stealth] (mask) -- (mul);
\draw[->, semithick, >=stealth] (mask.east) to[out=0,in=90] (cat.north);

\node[lbl, above=0.15cm of pad] {preserve\\64 memory\\embeddings};

\end{tikzpicture}
}
\caption{Local-additive conditioning for inpainting. Waveforms are encoded by a frozen SAME autoencoder into a latent sequence ($256 \times T$), then element-wise multiplied by a binary mask (1=keep, 0=inpaint). The masked latent and mask are concatenated along the channel dimension. Each transformer block projects this through an MLP with SiLU and adds the result to the residual stream. Left-padding leaves space to preserve the memory embeddings.}
\label{fig:local_additive_cond}
\end{figure}
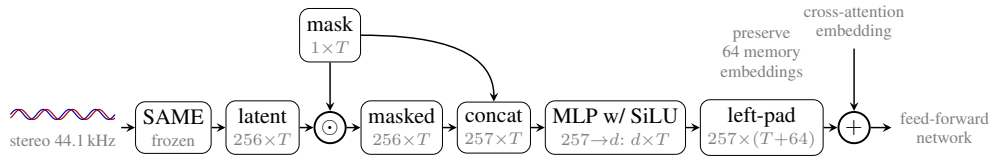

\paragraph{Differential attention.} Instead of a single set of queries
and keys, we build two pairs $(Q, K)$ and
$(Q', K')$ that share a common set of values~$V$. Two independent attention maps are computed
and their outputs are subtracted:
$\operatorname{Attn}(Q,\, K,\, V)-\operatorname{Attn}(Q',\, K',\, V)$, canceling the attention patterns common to both heads.
Both pairs undergo the same per-head QK-RMSNorm and, in the case of self-attention, partial RoPE is used. \texttt{medium} and \texttt{large} use differential attention~\cite{ye2025diffattn} in both self-attention and cross-attention, while \texttt{small} uses the standard multi-head attention.

\textbf{RMSNorm.} We use RMSNorm~\cite{liu2024deepseek} as a pre-normalization layer in transformer blocks. Given an input vector $\mathbf{x}$:
\begin{equation}
  \mathrm{RMSNorm}(\mathbf{x}) = \frac{\mathbf{x}}{\sqrt{\frac{1}{d} \lVert \mathbf{x} \rVert^2 + \epsilon}} \odot \boldsymbol{\gamma},
\end{equation}
where $\boldsymbol{\gamma} \in \mathbb{R}^d$ is a learnable scale parameter initialized to ones and $\epsilon = 10^{-5}$. Unlike LayerNorm, RMSNorm omits mean centering and the learnable bias, reducing computation while performing comparably in practice.

\textbf{QK-RMSNorm.} We apply per-head RMSNorm independently to Q and K after projection but before adding RoPE:
\begin{equation}
\hat{Q} = \text{RMSNorm}_q(Q), \quad \hat{K} = \text{RMSNorm}_k(K),
\end{equation}
where $\text{RMSNorm}_q$ and $\text{RMSNorm}_k$ have separate learnable scale parameters shared across all heads. This prevents the attention logits to grow unboundedly. QK-RMSNorm is applied in both self-attention and cross-attention~\cite{liu2024deepseek,henry2020query}.

\section{Training}\label{sec:training}

\begin{figure}[h]
\vspace{-1mm}
\centering
\resizebox{0.53\textwidth}{!}{%
\begin{tikzpicture}[
    >=stealth,
    stage/.style={rectangle, draw=gray!60, rounded corners=4pt, minimum height=0.8cm, minimum width=2.4cm, inner sep=5pt, font=\footnotesize, align=center},
    arr/.style={->, semithick, gray!70},
    annot/.style={font=\scriptsize, color=gray!70, align=center},
    brace/.style={decorate, decoration={brace, amplitude=4pt, raise=2pt}, gray!70, thick},
    bracelbl/.style={font=\scriptsize, color=gray!90, align=center},
]

\node[stage, fill=black!8] (rf) {Flow Matching\\Pre-Training};
\node[stage, fill=black!16, right=1.5cm of rf] (ode) {Distillation\\Warmup};
\node[stage, fill=black!24, right=1.5cm of ode] (arc) {Adversarial\\Post-Training};

\draw[arr] (rf) -- (ode);
\draw[arr] (ode) -- (arc);

\node[annot, below=4pt of rf] {velocity prediction\\MSE loss};
\node[annot, below=4pt of ode] {few-step ODE\\distillation};
\node[annot, below=4pt of arc] {reward-based\\refinement};

\draw[brace] (rf.north west) -- (rf.north east)
    node[bracelbl, midway, above=8pt] {base model};
\draw[brace] (ode.north west) -- (arc.north east)
    node[bracelbl, midway, above=8pt] {post-trained model};

\end{tikzpicture}
}
\vspace{-1mm}
\caption{Stable Audio 3 training pipeline.}
\label{fig:training_pipeline}
\end{figure}
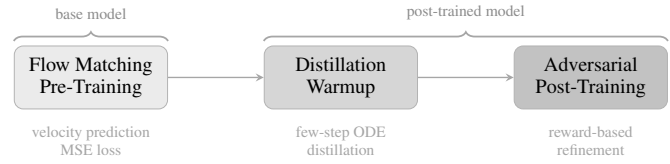

Stable Audio 3 models allow variable-length generation and are trained following a multi-stage pipeline (Figure~\ref{fig:training_pipeline}). The first stage trains the diffusion transformer ({base model}) with flow matching. Later stages (distillation warmup, adversarial post-training) refine the model for improved speed and sample quality ({post-trained model}). All stages operate on pre-encoded SAME latents encoded offline, and use the variable-length training schema below.

First, we train a flow matching model that learns a velocity field $v_\theta(x_t, t)$ defining an ordinary differential equation~(ODE) transporting noise $\epsilon$ to data $x_0$. At inference, this ODE is solved numerically over many $t$ steps (50--100).

Second, we perform a distillation warmup that repurposes the model as a one-step denoiser. Given any intermediate state $x_t$ sampled along the teacher's ODE trajectory, the student (same architecture as the teacher) learns to predict the trajectory's endpoint $\hat{x}_0$ (generation) directly, trained with an MSE loss. This effectively straightens the learned flow (collapsing the multi-step ODE solve into a single function evaluation $x_t \to \hat{x}_0$ for every $t$) but the MSE objective causes the student to regress toward the conditional mean $\mathbb{E}[\hat{x}_0 \mid x_t]$, producing outputs that lack fine-grained detail.

Third, adversarial post-training replaces the teacher signal with a relativistic adversarial setup that directly compares the student's one-step predictions  $x_t \to \hat{x}_0$ against real data ${x}_0$. This shifts the student's mapping from approximating the conditional mean toward sampling from the true data distribution $p(x_0 \mid x_t)$, recovering the perceptual sharpness that MSE distillation smooths over. Crucially, this stage discards the teacher entirely, allowing the student to surpass the teacher's quality ceiling by optimizing directly against real data $x_0$. 

While the resulting adversarially trained model can generate audio in a single forward pass, the mapping from pure noise $\epsilon \to \hat{x}_0$ in one step remains challenging (Section \ref{sec:eval_arc}). In Section~\ref{sec:inference}, we describe how ping-pong sampling alleviates this by decomposing the single large step into multiple smaller ones. At each iteration, the model produces a denoised estimate $\hat{x}_0$, which is then renoised with new noise at a reduced level before the next denoising step. This iterative denoise-then-renoise schedule allows the model to progressively refine its output, correcting errors from earlier steps while leveraging the one-step denoising $x_t \to \hat{x}_0$ capability learned during adversarial post-training.

\subsection{Variable-Length Training}\label{subsec:varlen}

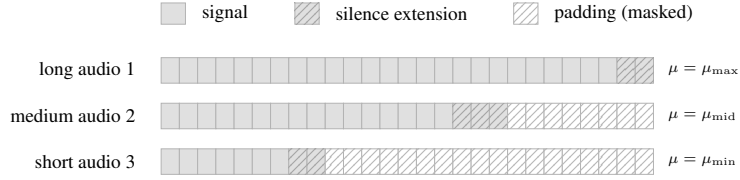
\begin{figure}[h]
\centering
\resizebox{0.6\textwidth}{!}{%
\begin{tikzpicture}[font=\small]

\def\cw{0.30}
\def\ch{0.42}
\def\rowgap{0.75}
\def\ntok{27}

\def\yrone{0}
\def\yrtwo{-1*\rowgap}
\def\yrthree{-2*\rowgap}

\fill[black!12, draw=gray!60, line width=0.4pt] (0,0.55) rectangle (0.40,0.90);
\node[anchor=west, font=\footnotesize] at (0.55,0.725) {signal};

\fill[black!12, draw=gray!60, line width=0.4pt] (2.2,0.55) rectangle (2.60,0.90);
\fill[pattern=north east lines, pattern color=black!35] (2.2,0.55) rectangle (2.60,0.90);
\node[anchor=west, font=\footnotesize] at (2.75,0.725) {silence extension};

\fill[white, draw=gray!60, line width=0.4pt] (5.8,0.55) rectangle (6.20,0.90);
\fill[pattern=north east lines, pattern color=gray!55] (5.8,0.55) rectangle (6.20,0.90);
\node[anchor=west, font=\footnotesize] at (6.35,0.725) {padding (masked)};

\node[anchor=east, font=\footnotesize] at (-0.3,\yrone-\ch/2) {long audio 1};
\node[anchor=east, font=\footnotesize] at (-0.3,\yrtwo-\ch/2) {medium audio 2};
\node[anchor=east, font=\footnotesize] at (-0.3,\yrthree-\ch/2) {short audio 3};

\foreach \i in {0,...,24} {
  \fill[black!12, draw=gray!60, line width=0.4pt]
    (\i*\cw,\yrone) rectangle (\i*\cw+\cw,\yrone-\ch);
}
\foreach \i in {25,26} {
  \fill[black!12, draw=gray!60, line width=0.4pt]
    (\i*\cw,\yrone) rectangle (\i*\cw+\cw,\yrone-\ch);
  \fill[pattern=north east lines, pattern color=black!35]
    (\i*\cw,\yrone) rectangle (\i*\cw+\cw,\yrone-\ch);
}

\foreach \i in {0,...,15} {
  \fill[black!12, draw=gray!60, line width=0.4pt]
    (\i*\cw,\yrtwo) rectangle (\i*\cw+\cw,\yrtwo-\ch);
}
\foreach \i in {16,17,18} {
  \fill[black!12, draw=gray!60, line width=0.4pt]
    (\i*\cw,\yrtwo) rectangle (\i*\cw+\cw,\yrtwo-\ch);
  \fill[pattern=north east lines, pattern color=black!35]
    (\i*\cw,\yrtwo) rectangle (\i*\cw+\cw,\yrtwo-\ch);
}
\foreach \i in {19,...,26} {
  \fill[white, draw=gray!60, line width=0.4pt]
    (\i*\cw,\yrtwo) rectangle (\i*\cw+\cw,\yrtwo-\ch);
  \fill[pattern=north east lines, pattern color=gray!55]
    (\i*\cw,\yrtwo) rectangle (\i*\cw+\cw,\yrtwo-\ch);
}

\foreach \i in {0,...,6} {
  \fill[black!12, draw=gray!60, line width=0.4pt]
    (\i*\cw,\yrthree) rectangle (\i*\cw+\cw,\yrthree-\ch);
}
\foreach \i in {7,8} {
  \fill[black!12, draw=gray!60, line width=0.4pt]
    (\i*\cw,\yrthree) rectangle (\i*\cw+\cw,\yrthree-\ch);
  \fill[pattern=north east lines, pattern color=black!35]
    (\i*\cw,\yrthree) rectangle (\i*\cw+\cw,\yrthree-\ch);
}
\foreach \i in {9,...,26} {
  \fill[white, draw=gray!60, line width=0.4pt]
    (\i*\cw,\yrthree) rectangle (\i*\cw+\cw,\yrthree-\ch);
  \fill[pattern=north east lines, pattern color=gray!55]
    (\i*\cw,\yrthree) rectangle (\i*\cw+\cw,\yrthree-\ch);
}

\node[font=\scriptsize, anchor=west] at (27.4*\cw,\yrone-\ch/2) {$\mu = \mu_{\max}$};
\node[font=\scriptsize, anchor=west] at (27.4*\cw,\yrtwo-\ch/2) {$\mu = \mu_{\mathrm{mid}}$};
\node[font=\scriptsize, anchor=west] at (27.4*\cw,\yrthree-\ch/2) {$\mu = \mu_{\min}$};

\end{tikzpicture}
}
\vspace{1mm}
\caption{Variable-length training. A batch contains sequences of different lengths, padded to a common (variable) size. Padding embeddings are excluded (masked) from the loss. Each audio receives a length-dependent timestep shift~($\mu$), with longer sequences shifted toward higher noise levels. The signal is randomly extended with silence.}
\label{fig:variable_length}
\end{figure}

Previous latent diffusion models in the audio domain operate on fixed-length sequences, padding shorter audio with silence to match the maximum training
length~\cite{evans2024longform,evans2024stableaudio}. Generating a short audio clip thus requires inference at full length, with most of the computation spent producing silence, since inference on shorter sequences than it was trained on leads to output degradation (Section \ref{sec:eval_varlen_music}). This effectively ties inference cost to the chosen maximum length, whilst many practical usecases need audio much shorter than that maximum. Stable Audio 3, instead, natively supports variable-length generation. During training, where batching requires uniform sequence lengths for efficiency, it relies on the following mechanisms: variable-length attention and masked loss computation, per-element timestep shifts, and silence augmentation. Figure~\ref{fig:variable_length} illustrates these mechanisms on a batch of three sequences.

\paragraph{Variable-length attention and masked loss.}
Sequences shorter than the batch maximum length are right-padded in latent space. Padding embeddings are excluded (masked) from both self-attention and feed-forward using variable-length flash attention~\cite{dao2023flashattention2}. Since padding positions are excluded (masked) from attention, their outputs are uninformative and the loss is computed only over valid signal positions (using a mask over the loss).
Cross-attention is not masked. Memory embeddings participate in all attention layers without masking, but are removed before any loss computation. Right-padded positions are also excluded (masked) from the adversarial loss.

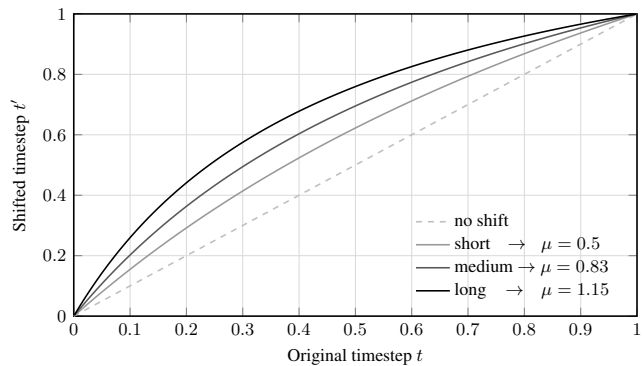
\begin{figure}[h]
\centering
\resizebox{0.52\textwidth}{!}{%
\begin{tikzpicture}
\begin{axis}[
    width=0.7\columnwidth,
    height=0.42\columnwidth,
    xlabel={Original timestep $t$},
    ylabel={Shifted timestep $t'$},
    xmin=0, xmax=1,
    ymin=0, ymax=1,
    grid=major,
    grid style={gray!30},
    legend pos=south east,
    legend style={
        font=\small,
        draw=none,
        fill=none,
        legend cell align=left,
        /tikz/every even column/.append style={column sep=4pt},
    },
    every axis label/.style={font=\small},
    every tick label/.style={font=\small},
    samples=100,
    domain=0.001:0.999,
    no markers,
    thick,
]

\addplot[gray!50, dashed, thick] coordinates {(0,0) (1,1)};
\addlegendentry{no shift}

\addplot[black!40, thick] {1 - exp(-0.5) / (exp(-0.5) + x/(1-x))};
\addlegendentry{short \hspace{1.5mm} $\rightarrow$ \hspace{1.1mm} $\mu=0.5$}

\addplot[black!65, thick] {1 - exp(-0.825) / (exp(-0.825) + x/(1-x))};
\addlegendentry{medium $\rightarrow$ $\mu=0.83$}

\addplot[black, thick] {1 - exp(-1.15) / (exp(-1.15) + x/(1-x))};
\addlegendentry{long \hspace{1.8mm} $\rightarrow$ \hspace{1.5mm} $\mu=1.15$}

\end{axis}
\end{tikzpicture}
}
\vspace{-1mm}
\caption{Effect of the per-element timestep shift on the timestep mapping. For short audios ($\mu_{\min}$$=$0.5), the shift is mild. For long audios ($\mu_{\max}$$=$1.15), timesteps are pushed substantially toward higher noise levels.}
\label{fig:dist_shift}
\end{figure}

\paragraph{Per-element timestep shifts.}
The noise schedule is adapted per sample based on its (unpadded) length.
Note that long sequences are harder to corrupt because of correlation across sequence elements.
Hence, when noise is added independently to each element of a sequence, longer sequences retain more recoverable structure at a given noise level due to redundancy between neighbouring elements~\cite{hoogeboom2023simple,chen2023noise}. This means that a fixed noise schedule can under-noise long sequences relative to short ones, biasing the model toward learning to denoise at insufficiently high noise levels for long inputs.
To compensate, the timestep distribution is shifted toward higher noise levels for longer sequences. The shift pushes longer sequences toward noisier timesteps (Figure~\ref{fig:dist_shift}), giving the model more training budget in the high-noise regime. The proposed shift uses the logistic form proposed by Esser et al.~\cite{esser2024sd3}. Given a parameter $\mu$ that {interpolates between} $\mu_{\min}$$=$0.5 and $\mu_{\max}$$=$1.15 as a function of the sequence length, the shifted timestep is:
\begin{equation}
\label{eq:shift_varlen}
t' = 1 - \frac{e^{-\mu}}{e^{-\mu} + \frac{t}{1-t}}.
\end{equation}

\vspace{-4mm}

\paragraph{Silence augmentation.} To improve robustness and break the direct correspondence between duration conditioning and signal length, the signal region is randomly extended with silence embeddings with a length drawn from an exponential distribution (on average 4 sec of silence extension). The padding is filled with a pre-computed silence latent (obtained by encoding a zero-valued waveform) so that the model encounters realistic silence representations. This teaches the model to terminate audio cleanly with natural silence rather than abrupt cutoffs or artifacts.

\subsection{Flow Matching Pre-Training}\label{subsec:flow_matching}

The initial training stage uses a flow matching objective~\cite{liu2022flow,lipman2023flow}.
Given a data sample $x_0$ (SAME latents) and noise $\epsilon \sim \mathcal{N}(0, I)$, the noised input at timestep $t \in [0, 1]$ is the linear interpolation:
\begin{equation}
x_t = (1 - t)\, x_0 + t\, \epsilon,
\end{equation}
and the model is trained to predict the velocity $v = \epsilon - x_0$ via a mean squared loss with inpainting masks.

\paragraph{Inpainting training.} All models are trained jointly for generation and inpainting. At each training step, a random binary mask is sampled per example, where $m$$=$1 denotes positions to keep the audio and $m$$=$0 positions to inpaint. 
Three mask types are drawn with probabilities: \emph{full mask} with all zeros is equivalent to unconditional generation (probability 80\%), \emph{random segments} where 1 to 10 segments are masked out for inpainting (probability 10\%), and \emph{causal mask} where a random prefix is kept and the remainder is masked for continuation (probability 10\%). Inpainting mask examples are in Figure \ref{fig:inpainting_masks}. The mask and the element-wise product of the clean latent with the mask are concatenated along the channel dimension and provided to the model as local-additive conditioning. The model is trained to predict the velocity and the loss is split into two independently averaged terms: a generation loss over the inpainted embeddings ($m$$=$0) and a context preservation loss over the kept audio ($m$$=$1).
\begin{equation}
\frac{1}{N_{\text{gen}}}\!\sum_{i:\,m=0}\!\left\|\hat{v}_\theta(x_t, t, c)_i - (\epsilon - x_0)_i\right\|^2 \;+\; \frac{1}{N_{\text{ctx}}}\!\sum_{j:\,m=1}\!\left\|\hat{v}_\theta(x_t, t, c)_j - (\epsilon - x_0)_j\right\|^2\,,
\end{equation}
where $\hat{v}_\theta$ is the predicted velocity, $c$ is any conditioning signal, and $N_{\text{gen}}$, $N_{\text{ctx}}$ are the number of inpainted ($m\!=\!0$) and context ($m\!=\!1$) embeddings, respectively.

\paragraph{Minibatch optimal transport coupling.} It is used to find a permutation of noise samples that minimises the squared $L_2$ transport cost within each minibatch, computed via Sinkhorn iterations on GPU \cite{liu2022flow,cuturi2013sinkhorn}. Concretely, given a batch of $B$ data samples and $B$ independently drawn noise vectors, the algorithm reassigns which noise vector is paired with which data sample. Standard flow matching pairs each data sample $x_0$ with an independently drawn noise sample $\epsilon$. When these happen to be far apart, the resulting transport path is long and may cross paths from other pairs. By solving an approximate assignment problem within each minibatch, optimal transport coupling pairs each $x_0$ with the closest available $\epsilon$, producing shorter, straighter, and less entangled trajectories. This straightens the velocity field that the model learns, improving both training and sampling~\cite{lan2024high}.

\paragraph{Timestep sampling.}
Timesteps are drawn from a truncated logit-normal distribution \cite{flux2024,esser2024sd3}. Samples from a logit-normal distribution ($t = \sigma(z),\; z \sim \mathcal{N}(0,1)$) are truncated at $t = 0.075$ and rescaled to $[0, 1]$.
This removes very-low-noise timesteps and concentrates training budget on intermediate-high noise levels. Remember that due to our variable-length training schema, the sampled $t$ are individually shifted $t'$ based on Equation \ref{eq:shift_varlen}.

\subsection{Distillation Warmup}\label{subsec:ode_warmup}

The distillation warmup bridges the gap between the many-step flow matching model and the one-step regime required by adversarial post-training, providing a smoother initialisation than directly addressing the adversarial objective.

Before adversarial training, the flow matching model is refined through a distillation warmup stage of 10k steps \cite{luhman2021distillation}. A frozen copy of the pre-trained flow matching model serves as a teacher, and the student is also initialized with the pre-trained flow matching model. The teacher generates a multi-step ODE trajectory (15 DPM++ steps with CFG set to 5) from random noise $\epsilon$, and the intermediate states $(x_t, t)$ along with the final denoised output $\hat{x}_0$ are cached. The teacher trajectory is refreshed periodically (every 4 iterations) to balance compute cost with target diversity. The student is then trained to match the teacher's endpoint $\hat{x}_0$ in a single step: given a randomly selected intermediate state $x_t$ from the cached trajectory, the student predicts a velocity $v_\theta(x_t, t, c)$ and produces a one-step Euler estimate $\hat{x}_{0,\theta} = x_t - t \, v_\theta(x_t, t, c)$.
The loss is the MSE between $\hat{x}_{0,\theta}$ and the teacher's denoised output $\hat{x}_0$:
$$\mathcal{L} = \| (x_t - t\, v_\theta(x_t, t, c)) - \hat{x}_0 \|^2.$$

Note that our technique is related to ReFlow~\cite{liu2022flow}, which straightens the transport paths of a trained flow model. In standard flow matching, each data sample $x_0$ is paired with an independently drawn noise sample $\epsilon$, resulting in random couplings that may produce long, crossing transport paths. ReFlow addresses this by generating coupled endpoints~$(\hat{x}_0, \epsilon)$. It samples noise $\epsilon$ and integrates the trained model's ODE to obtain the corresponding output $\hat{x}_0$. A new model is then trained to connect these coupled endpoints (with straighter paths that require fewer sampling steps, enabling faster inference). Although our approach also unrolls the teacher's ODE, it differs in a key respect to ReFlow. Rather than retraining a flow matching model (predict velocity) on these new endpoint pairs, our student learns to map any intermediate state $x_t$ along the teacher's trajectory directly to the final output $\hat{x}_0$ in a single step.

It is also related to one-step ReFlow~\cite{liu2022flow}, which further distills the reflowed model into a single-step flow matching model (velocity prediction) that maps $\epsilon \to \hat{x}_{0}$ with one Euler step. Our approach shares the same one-step generation approach, but bypasses the intermediate ReFlow training stage entirely. Further, our student learns to map any intermediate state $x_t$ along the teacher's ODE trajectory to the final output $\hat{x}_0$ based on the MSE loss (instead of velocity prediction). This is handy because during adversarial post-training our losses operate in signal ${x}_0$ space instead of vector field $v_\theta(x_t, t)$ space, but this can introduce regression-to-the-mean artifacts despite our teacher providing a unique $\hat{x}_0$ for each $x_t$. This motivates the adversarial post-training stage, where the discriminator's loss in ${x}_0$ space directly penalizes perceptual degradation and recovers the fine-grained structure that MSE distillation can smooth over.

Finally, our setup is also related to Consistency Distillation~\cite{song2023consistency} which trains a student model to map any point on the teacher's ODE trajectory to the endpoint $x_t \to \hat{x}_0$.
It relies on a {local-consistency} loss where the model's predictions at two adjacent steps ${x}_{t}$ and ${x}_{t+1}$ along the same ODE trajectory must be the same $f_\theta({x}_{t+1}, t+1) {\approx} f_{\theta^-}({x}_{t},t)$, where $\theta^-$ denotes an exponential moving average of the student weights. Yet, local consistency alone does not determine the value trajectory steps should map to. What anchors the local consistency chain to the actual endpoint $\hat{x}_0$ is a boundary condition at $t{=}0$. During training, the model is architecturally constrained to bypass $\hat{x}_0$ so that $f_\theta(\hat{x}_0, t{=}0){=}\hat{x}_0$ by construction. The consistency loss then chains everything together: the prediction at $t{=}0.1$ must match the prediction at $t{=}0$, which is hardwired to $\hat{x}_0$. The prediction at $t{=}0.2$ must match $t{=}0.1$, which already equals $\hat{x}_0$. This cascade propagates all the way to $t{=}1$, so that even from pure noise the model can predict $\hat{x}_0$. Our approach does not propagate information through a chain of local consistency losses anchored by a boundary condition. Instead, we regress directly to the teacher's endpoint $\hat{x}_0$ from any intermediate state $\hat{x}_t$. A trade-off of our MSE-based regression approach is that if the teacher's ODE is curved, our endpoint $\hat{x}_0$ estimates won't be accurate (which is why ping-pong sampling is required, see Section \ref{sec:inference}). Consistency Distillation sidesteps this by focusing on consistency across adjacent time-steps.

\subsection{Adversarial Post-Training}
\label{subsec:arc}

Adversarial post-training turns our pre-trained model into a one-step generator by supplanting the MSE-based conditional {mean} loss (of both flow matching and distillation warmup) with an adversarial loss. To that end, a discriminator evaluates the {realism} of denoised samples, providing distribution-level feedback that goes beyond the broad conditional {mean} loss.  As such, if the denoised output $x_t \to \hat{x}_0$ is sufficiently {real} and higher-quality, fewer sampling steps are required. Another key advantage of adversarial post-training over distillation methods is that it sidesteps reliance on the performance of the pre-trained (teacher) model.
The goal of adversarial post-training is to map $x_t \to \hat{x}_0$ where  ${x}_t = (1{-}t)\, {x_0} + t\,{\epsilon}$ and $\hat{x}_0$ is an estimate of $x_0$. A relativistic discriminator, operating in ${x}_0$ space, provides the training signal required to recover the details that MSE-based distillation smooths over, effectively trading regression-to-the-mean artifacts for perceptually sharp outputs while preserving the one-step capabilities of the distilled model.

We fine-tune the pre-trained model (flow matching plus distillation warmup) using adversarial post-training with three complementary losses: an adversarial relativistic loss~$\mathcal{L}_R$, a contrastive loss~$\mathcal{L}_C$, and a CLAP loss~$\mathcal{L}_{\text{CLAP}}$.

Training alternates between generator (the pre-trained one-step model) and discriminator updates:
\begin{align}
    \text{Generator:} \quad & \mathcal{L}_G = \mathcal{L}_R^{(G)} + \mathcal{L}_{\text{CLAP}}^{(G)}, \label{eq:gen-loss} \\
    \text{Discriminator:} \quad & \mathcal{L}_D = \mathcal{L}_R^{(D)} + \mathcal{L}_C^{(D)}. \label{eq:disc-loss}
\end{align}

$\mathcal{L}_R$ drives perceptual quality, $\mathcal{L}_C$ regularizes the discriminator to be semantically aligned, and $\mathcal{L}_{\text{CLAP}}$ gives the generator an explicit text-alignment signal such that the generator improves both audio fidelity and prompt alignment.

\paragraph{Generator architecture.}
It is the pre-trained model with flow matching and distillation warmup.
In principle, one could parameterize the network to output $\hat{x}_0$
directly. Yet, we retain the original velocity parameterization $v_\theta(x_t, t, c)$
from the base model and recover the clean estimate via one-step Euler sampling:
$\hat{x}_0 = x_t - t \, v_\theta(x_t, t, c)$.
Note this reparameterization also imposes a useful architectural constraint:
at $t{=}0$ the model must output $\hat{x}_0 {=} x_0 - 0 \cdot v_\theta {=} x_0$.
As $t$ grows, the network's influence scales linearly with the noise level,
preventing it from making disproportionately large corrections at low noise
where the input is already close to clean.
Further, it preserves initialization quality: since the generator starts from
$v_\theta$, the initial outputs are already meaningful predictions, improving early training
stability.
The same reparameterization is used for the distillation warmup: maintaining the one-step Euler sampling reparametrization throughout post-training ensures a smooth transition from flow matching without any discontinuities.

\newpage

\paragraph{Discriminator architecture.}
The discriminator reuses the same architecture as the generator as a feature extractor,
initialized from the base model pretrained with flow matching (no distillation warmup).
It is a {fully-conditioned} discriminator that receives the text prompt, the duration conditioning, the inpainting mask and context, and a timestep $t_D$ (independent from $t$ used by the generator). All signals go through the same conditioning mechanisms (cross-attention, adaptive layer norm) as the generator. Having been pre-trained to process noisy data $x_t$ under arbitrary conditions, it already produces semantically rich intermediate representations without any additional training. Such intermediate representations are then processed by a convolutional head that produces frame-wise realness scores.

The discriminator operates at a noise level $t_D$ that is {independent}
of the generator's noise level $t$.
Specifically, after the generator produces its denoised estimate
$\hat{x}_0 = x_t - t \, v_\phi(x_t, t)$, we renoise it to a fresh
noise level for the discriminator:
\begin{equation}
    x_{t_D}^{\text{fake}} = (1 - t_D)\,\hat{x}_0 + t_D\,\epsilon'  \quad
    x_{t_D}^{\text{real}} = (1 - t_D)\,x_0 + t_D\,\epsilon'
    \label{eq:disc_noise}
\end{equation}
where $t_D$ evaluates samples at timesteps drawn from a logit-normal distribution ($t_D = \sigma(z),\; z \sim \mathcal{N}(0,1)$), focusing on intermediate noise levels, and
$\epsilon' \sim \mathcal{N}(0, I)$ is shared between real and fake.
The decoupling of $t$ and $t_D$ allows the discriminator to evaluate the generator's output at {multiple noise scales}, providing training signal about both global structure (high $t_D$) and fine detail (low $t_D$), regardless of which $t$ the generator was trained on in that iteration. Finally, by using the same noise $\epsilon'$ when constructing ${x}_{t_D}^{\text{real}}$ and ${x}_{t_D}^{\text{fake}}$, the noise component cancels in the relativistic comparison, ensuring the discriminator judges the quality difference between ${x}_0$ and $\hat{{x}}_0$ rather than incidental noise patterns.

\paragraph{Adversarial relativistic loss.} The generator is trained to minimize $D_\psi({x}_{t_D}^{\text{real}}, t_D, c) - D_\psi({x}_{t_D}^{\text{fake}}, t_D, c)$ and the discriminator is trained to maximize this margin.
Our loss relies on the $\text{softplus}(x) = \log(1 + e^x)$ as a smooth surrogate for $\max(0, x)$. Unlike a raw margin loss that would continue rewarding an already-winning player (with low $\mathcal{L}_R$), softplus saturates: when it is large and negative, $\text{softplus}(x) \approx 0$, gradients vanish gracefully. Conversely, when the argument is positive, $\text{softplus}(x) \approx x$ and the loss grows linearly, providing a strong corrective signal.
\begin{align}
\mathcal{L}_R^{(G)} &= \mathbb{E}\!\left[\,\text{softplus}\!\Big(D_\psi({x}_{t_D}^{\text{real}}, t_D, c) - D_\psi({x}_{t_D}^{\text{fake}}, t_D, c)\Big)\,\right] \label{eq:lr-gen} \\
\mathcal{L}_R^{(D)} &= \mathbb{E}\!\left[\,\text{softplus}\!\Big({-}\big(D_\psi({x}_{t_D}^{\text{real}}, t_D, c) - D_\psi({x}_{t_D}^{\text{fake}}, t_D, c)\big)\Big)\,\right]
\end{align}

The loss is calculated on {pairs} of real/generated data \cite{huang2024gan,jolicoeur2018relativistic}, such that the generator minimizes its detection \textit{relative} to its paired real sample with the same prompt. 
Thus, the generator wants every generated sample to be ``more real than its paired real sample", while the discriminator wants every real sample to be ``more real than its paired generated sample". {Critically, our pairs are always {highly} related due to our text-conditional task, where pairs of real/generated share the same prompt, thus providing a stronger gradient signal than relying on random pairings.}


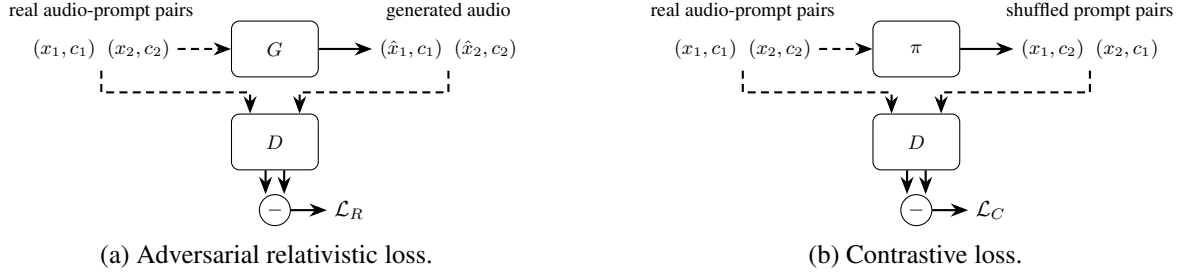
\begin{figure}[t]
\vspace{-4mm}
\centering
\tikzset{
  >=Stealth,
  arcblock/.style={draw, rounded corners=3pt, minimum height=0.8cm,
                   align=center, font=\small},
  gbox/.style    ={arcblock},
  dbox/.style    ={arcblock},
  pbox/.style    ={arcblock},
  diffcirc/.style={circle, draw, inner sep=0pt, minimum size=5mm,
                   font=\small, fill=white},
  darrow/.style  ={->, thick, densely dashed},
  sarrow/.style  ={->, thick},
}
\begin{minipage}[t]{0.48\textwidth}
\centering
\begin{tikzpicture}[scale=0.82, transform shape]

  \node[font=\small] at (-2.8,3.9) {real audio-prompt pairs};
  \node[font=\small] at (2.8,3.9) {generated audio};

  \node[font=\small] (rc1) at (-3.4,3.3) {$(x_1, c_1)$};
  \node[font=\small] (rc2) at (-2.2,3.3) {$(x_2, c_2)$};

  \node[gbox, minimum width=1.4cm, minimum height=0.9cm] (G) at (0,3.3)
       {$G$};

  \node[font=\small] (g1) at (2.2,3.3) {$(\hat{x}_1, c_1)$};
  \node[font=\small] (g2) at (3.4,3.3) {$(\hat{x}_2, c_2)$};

  \draw[darrow] (rc2.east) -- (G.west);
  \draw[sarrow] (G.east) -- (g1.west);

  \node[dbox, minimum width=1.4cm, minimum height=0.9cm]
      (D) at (0,1.8) {$D$};

  \draw[darrow]
      (-2.8,2.95) -- (-2.8,2.6) -| ([xshift=-4mm]D.north);
  \draw[darrow]
      (2.8,2.95) -- (2.8,2.6) -| ([xshift=4mm]D.north);

  \node[diffcirc] (minus) at (0,0.7) {$-$};

  \draw[sarrow]
      ([xshift=-1.5mm]D.south) -- ([xshift=-1.5mm]minus.north);
  \draw[sarrow]
      ([xshift=1.5mm]D.south) -- ([xshift=1.5mm]minus.north);

  \node[font=\normalsize\bfseries] (loss) at (1.2,0.7) {$\mathcal{L}_{R}$};
  \draw[sarrow] (minus.east) -- (loss.west);

\end{tikzpicture}
\vspace{1mm}

{(a)} Adversarial relativistic loss.
\end{minipage}
\hfill
\begin{minipage}[t]{0.48\textwidth}
\centering
\begin{tikzpicture}[scale=0.82, transform shape]

  \node[font=\small] at (-2.8,3.9) {real audio-prompt pairs};
  \node[font=\small] at (2.8,3.9) {shuffled prompt pairs};

  \node[font=\small] (rc1) at (-3.4,3.3) {$(x_1, c_1)$};
  \node[font=\small] (rc2) at (-2.2,3.3) {$(x_2, c_2)$};

  \node[pbox, minimum width=1.4cm, minimum height=0.9cm] (P) at (0,3.3)
       {$\pi$};

  \node[font=\small] (s1) at (2.2,3.3) {$(x_1, c_2)$};
  \node[font=\small] (s2) at (3.4,3.3) {$(x_2, c_1)$};

  \draw[darrow] (rc2.east) -- (P.west);
  \draw[sarrow] (P.east) -- (s1.west);

  \node[dbox, minimum width=1.4cm, minimum height=0.9cm]
      (D) at (0,1.8) {$D$};

  \draw[darrow]
      (-2.8,2.95) -- (-2.8,2.6) -| ([xshift=-4mm]D.north);
  \draw[darrow]
      (2.8,2.95) -- (2.8,2.6) -| ([xshift=4mm]D.north);

  \node[diffcirc] (minus) at (0,0.7) {$-$};

  \draw[sarrow]
      ([xshift=-1.5mm]D.south) -- ([xshift=-1.5mm]minus.north);
  \draw[sarrow]
      ([xshift=1.5mm]D.south) -- ([xshift=1.5mm]minus.north);

  \node[font=\normalsize\bfseries] (loss) at (1.2,0.7) {$\mathcal{L}_{C}$};
  \draw[sarrow] (minus.east) -- (loss.west);

\end{tikzpicture}
\vspace{1mm}

{(b)} Contrastive loss.
\end{minipage}
\caption{Adversarial Post-Training.
{(a)}~Pairs of generated and real samples (with the same text prompts) are passed into the discriminator (with additive noise), where the generator and discriminator are trained to minimize and maximize (respectively) the difference between fake and real outputs.
{(b)}~The discriminator is also trained to maximize the difference between audios with correct and incorrect (shuffled) prompts. Dashed lines denote noise injection.}
\label{fig:arc-losses}
\end{figure}


\paragraph{Contrastive loss.}
To prevent the discriminator from relying on audio-only artifacts and ignoring the text conditioning, we add a contrastive term. Given a batch of real audio-prompt pairs, we create negative examples by cyclically shifting (random) the prompts across the batch. The discriminator is then trained to distinguish correctly paired from incorrectly paired audio-prompt combinations, using the same relativistic objective that maximizes their difference:
\begin{equation}
    \mathcal{L}_C^{(D)} = \mathbb{E}\!\left[\,\text{softplus}\!\Big({-}\big(D_\psi(\mathbf{x}_t \mid c_{\text{correct}}) - D_\psi(\mathbf{x}_t \mid c_{\text{shuffled}})\big)\Big)\,\right].
    \label{eq:lc}
\end{equation}

This loss can be viewed as a contrastive loss~\cite{gutmann2010noise} where the discriminator is mapping correct audio-text pairs closer than mismatched pairs.
Note that this is only a loss for the {discriminator}, as this encourages it to understand the alignment between prompts and noisy inputs, and prevents the model from focusing on easier (\textit{e.g.}, high-frequency~\cite{Novack2025Presto}) audio features. This forces the discriminator to understand audio-text alignment, not just audio quality.

\paragraph{CLAP loss.}
A frozen CLAP model provides direct semantic guidance to the generator. The generator's denoised output $\hat{\mathbf{x}}_0$ and the text prompt are encoded, and we minimize the squared geodesic distance on the unit hypersphere:
\begin{equation}
    \mathcal{L}_{\text{CLAP}}^{(G)} = 2\,\arcsin^2\!\!\left(\frac{\lVert{e}_{\text{text}} - {e}_{{\hat{x}_0}} \rVert_2}{2}\right),
    \label{eq:lclap}
\end{equation}

where ${{e}}_{\text{text}}$ and ${{e}}_{\text{audio}}$ are $\ell_2$-normalized embeddings from the CLAP text and audio encoders, respectively. Since decoding latents (to waveform) at each training step would be expensive, we train a CLAP model that operates directly on SAME embeddings, avoiding the need for waveform synthesis during adversarial post-training.
We adopt the squared geodesic distance on the unit hypersphere as our CLAP alignment objective, since unlike cosine distance (whose gradient vanishes at both small and large angular separations) it provides a gradient magnitude proportional to the angle between embeddings across the full range. This ensures a consistent learning signal, particularly in early stages when text and audio embeddings may be misaligned. 
Hence, the CLAP loss provides a semantic anchor that prevents mode collapse during adversarial training by encouraging the generator to remain aligned with the prompt.

\subsection{Training Implementation Details}

\paragraph{Training data.} \texttt{medium} and \texttt{large} are trained on a combination of licensed audio from AudioSparx and Creative Commons recordings from Freesound. The AudioSparx portion (806,284 audios) contains music tracks, instruments, and sound effects with text metadata. The Freesound portion consists of recordings licensed under CC-0, CC-BY, or CCSampling+. To ensure no copyrighted content was present in the Freesound data, music recordings were identified using the PANNs~\cite{kong2020panns} tagger. We flagged audio that activated music-related tags for at least 30s (threshold of 0.15), that was sent to a trusted content detection company to verify the absence of copyrighted material. All identified copyrighted content was removed. After filtering, the Freesound part includes 266,324 CC-0, 194,840 CC-BY, and 11,454 CC-Sampling+ recordings\footnote{The same subset of Freesound audio we used to train Stable Audio Open: \url{https://info.stability.ai/attributions}.}. 
All \texttt{small} models are initially pre-trained on a mixture of AudioSparx and Freesound. But for the final stage of pre-training, distillation warmup, and post-training, we use AudioSparx for \texttt{small-music} and a higher-quality subset of Freesound for \texttt{small-sfx}. As a result, note that \texttt{medium} and \texttt{large} models are able to handle both music and sound effect generation within a single unified model. However, we find that for \texttt{small} models the inclusion of sound effects data degrades musical coherence. By isolating the sound effects subset into \texttt{small-sfx}, we mitigate this interference and obtain improved perceptual quality in both domains.

\paragraph{Prompt preparation.} For AudioSparx tracks, metadata includes natural-language descriptions, titles, keywords, and domain-specific metadata such as BPM, genre, moods, or instruments. For Freesound recordings, metadata includes titles, tags, and natural-language descriptions. During training, prompts are constructed by concatenating a random subset of available metadata fields. For AudioSparx,
with 50\% probability, the metadata field identifier is prepended (\textit{Instruments: Guitar, Drums, Bass Guitar, Moods: Uplifting}) or omitted. For Freesound, with 50\% probability, we use an LLM-rewritten caption of the audio or we construct a prompt from the title, shuffled tags, and description. In both cases (AudioSparx and Freesound), with 50\% probability, all candidate fields are shuffled or randomly subsampled. Finally, fields are joined with commas and, with 50\% probability, the entire prompt is lowercased.

\paragraph{Flow matching pre-training.} Classifier-free guidance (CFG) is enabled at training time by randomly replacing the conditioning embeddings (text and timing) with zero vectors with probability $p{=}0.1$, allowing guidance at inference.

\paragraph{Discriminator architecture.} Latent features are extracted from layer 14 of the discriminator's diffusion transformer and passed through a convolutional head consisting of: an input convolution, 4 residual blocks, each containing 2 convolutions with GroupNorm and LeakyReLU with a skip connection, and a final scoring module with 2 convolutions reducing to a single channel, producing per-embedding scores. All convolutional kernels are of size 3.

\paragraph{Timestep sampling.}
We use different distributions for each training stage. In all cases, the sampled timestep $t$ is subsequently shifted based on the effective (unpadded) sequence length, yielding $t'$ as defined in Equation~\ref{eq:shift_varlen}.

\begin{itemize}
    \item {Flow matching:} Timesteps are drawn from a truncated logit-normal distribution~\cite{esser2024sd3, flux2024}. Samples from a logit-normal ($t = \sigma(z),\; z \sim \mathcal{N}(0,1)$) are truncated at $t = 0.075$ and rescaled to $[0, 1]$. This removes near-clean timesteps where the task is trivial and concentrates training on intermediate-to-high noise levels.

    \item {Distillation warmup:} The teacher generates reference trajectories using DPM++ with 15 steps and CFG 5.0. The teacher's schedule uses a log-SNR schedule \emph{not} shifted based on sequence length, see Section \ref{sec:inference}.  

    \item Generator: the same setup (truncated logit-normal distribution) as described above for flow matching.

    \item {Discriminator:} Both real and generated samples are noised to the {same} $t_D$ (logit-normal) with {shared} noise $\epsilon$.
    
\end{itemize}

\paragraph{Optimizer.}
We use the Muon+AdamW hybrid optimizer~\cite{jordan2024muon}. Muon (momentum 0.95, learning rate $10^{-5}$) is applied to attention QKV projections and feed-forward network projections, while AdamW (learning rate $10^{-6}$, $\beta = (0.9, 0.95)$, weight decay $0.01$) handles all remaining parameters. The same optimizer configuration is used for both generator and discriminator. The learning rate follows an inverse power-law schedule ($\gamma{=}10^6$, power $0.5$).

\paragraph{Exponential Moving Average (EMA).}
We maintain an EMA of the generator weights ($\beta{=}0.9995$, power-law warmup with exponent $0.75$) throughout training and inference. EMA is not applied to the discriminator.

\section{Inference}
\label{sec:inference}

With adversarial post-training, our model is fine-tuned to directly estimate clean outputs $\hat{x}_0$ from noisy inputs $x_t$ at arbitrary noise levels $t$.
As such, the resulting adversarially post-trained model can generate audio in a single pass. Yet, one step mappings $\epsilon \to \hat{x}_0$ remain challenging (Section \ref{sec:eval_arc}). To adjust for that, we employ ping-pong sampling that addresses this by decomposing the single large step into multiple smaller ones. At each iteration, the model produces a denoised estimate $\hat{x}_0$, which is then renoised with new noise at a reduced level before the next denoising step. This iterative denoise-then-renoise schedule allows the model to progressively refine its output, correcting errors from earlier steps while leveraging the one-step denoising $x_t \to \hat{x}_0$ capability learned during adversarial post-training.

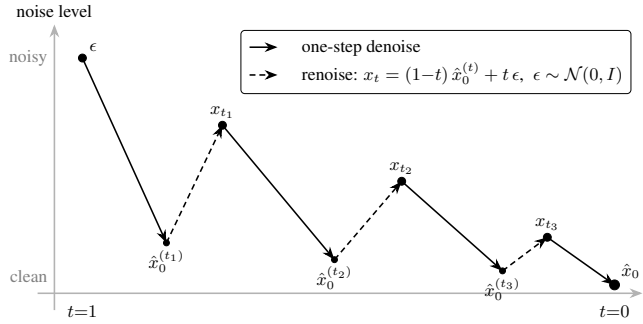
\begin{figure}[h]
\centering
\resizebox{0.52\textwidth}{!}{%
\begin{tikzpicture}[>=Stealth, scale=1.0]

  \draw[->, thick, gray!60] (-0.3,0) -- (10.5,0);
  \draw[->, thick, gray!60] (0,-0.3) -- (0,4.8) 
    node[above, font=\small, black] {noise level};

  \node[below, font=\small] at (0.5,-0.1) {$t{=}1$};
  \node[below, font=\small] at (10,-0.1) {$t{=}0$};

  \node[left, font=\footnotesize, gray] at (0,4.2) {noisy};
  \node[left, font=\footnotesize, gray] at (0,0.3) {clean};

  \filldraw[black] (0.5,4.2) circle (2pt) 
    node[above right, font=\footnotesize] {$\epsilon$};

  \draw[->, thick] (0.5,4.2) -- (2.0,0.9);
  \filldraw[black] (2.0,0.9) circle (1.5pt)
    node[below, font=\footnotesize] {$\hat{x}_0^{(t_1)}$};

  \draw[->, thick, densely dashed] (2.0,0.9) -- (3.0,3.0);
  \filldraw[black] (3.0,3.0) circle (2pt)
    node[above, font=\footnotesize] {$x_{t_1}$};

  \draw[->, thick] (3.0,3.0) -- (5.0,0.6);
  \filldraw[black] (5.0,0.6) circle (1.5pt)
    node[below, font=\footnotesize] {$\hat{x}_0^{(t_2)}$};

  \draw[->, thick, densely dashed] (5.0,0.6) -- (6.2,2.0);
  \filldraw[black] (6.2,2.0) circle (2pt)
    node[above, font=\footnotesize] {$x_{t_2}$};

  \draw[->, thick] (6.2,2.0) -- (8.0,0.4);
  \filldraw[black] (8.0,0.4) circle (1.5pt)
    node[below, font=\footnotesize] {$\hat{x}_0^{(t_3)}$};

  \draw[->, thick, densely dashed] (8.0,0.4) -- (8.8,1.0);
  \filldraw[black] (8.8,1.0) circle (2pt)
    node[above, font=\footnotesize] {$x_{t_3}$};

  \draw[->, thick] (8.8,1.0) -- (10.0,0.15);
  \filldraw[black] (10.0,0.15) circle (2.5pt)
    node[above right, font=\footnotesize] {$\hat{x}_0$};

  \node[draw, rounded corners=2pt, fill=white, inner sep=4pt, 
        font=\footnotesize, anchor=north east] at (10.3,4.7) {
    \begin{tabular}{@{}cl@{}}
      \tikz\draw[->, thick] (0,0) -- (0.5,0); & 
        one-step denoise \\[2pt]
      \tikz\draw[->, thick, densely dashed] (0,0) -- (0.5,0); & 
        renoise: $x_{t} = (1{-}t)\,\hat{x}_0^{(t)} + t\,\epsilon$,\; $\epsilon \sim \mathcal{N}(0, I)$
    \end{tabular}
  };

\end{tikzpicture}
}
\caption{Ping-pong sampling. Starting from pure noise $\epsilon \sim \mathcal{N}(0,I)$ at $t{=}1$, each step alternates between denoising to an $\hat{x}_0^{(t)}$ estimate (solid arrows) and stochastically re-noising with new noise at a lower timestep (dashed arrows). The re-noising magnitude decreases as $t \to 0$, producing a zigzag trajectory that converges to the data manifold.}
\label{fig:pingpong}
\end{figure}

Note that ping-pong sampling is inherently self-correcting. If early steps (where $x_t \to \hat{x}_0$ is more difficult) produce an inaccurate estimate, the subsequent re-noising step yields a new state that can correct the previous (difficult) estimate. 
In contrast, ODE solvers propagate errors forward, \textit{e.g.}: a bad Euler step displaces $x_t$ from the actual best trajectory, and all subsequent steps integrate from the wrong starting point.
This self-correcting property of ping-pong sampling means that our approach is tolerant to moderate deviations and errors throughout inference steps.

Instead of spacing timesteps linearly in $t \in [0,1]$, we construct a schedule that is uniform in logSNR space.
We define $N{+}1$ ($N{=}8$) equally-spaced logSNR steps $\lambda_0, \dots, \lambda_N$ in the interval
$[\lambda_{\min}, \lambda_{\max}] = [-6.2,\; 2.0]$ and recover\footnote{In flow matching $x_t = (1{-}t)\,x_0 + t\,\epsilon$, so the logSNR is
$\lambda(t) = \log\!\bigl(\tfrac{1-t}{t}\bigr)$.
Inverting this relationship gives $t = \tfrac{1}{1 + e^{\lambda}} = \sigma(-\lambda)$.} timesteps via
$t_i = \sigma(-\lambda_i)$.
This follows from perceptual salience being approximately uniform in logSNR space~\cite{kingma2021variational}. We found that 8 sampling steps provide a favorable trade-off between inference efficiency and generation quality.

During training, we employ a length-dependent timestep shift. Yet, at inference, we use a logSNR-uniform schedule that is \emph{not} length-dependent (the same schedule is used regardless of the requested duration).
While this introduces a train--inference mismatch,  this setup works well in practice. Note that during training a large amount of timesteps are being sampled (coverage at higher-noises matters), while inference takes only 8 steps (placement matters).

Crucially, the generation length (input noise $\epsilon$ length) is variable, see Figure \ref{fig:variable-length}.
Given a requested duration of $d$ seconds, we allocate a latent sequence of
$L = \lceil (d + d_{\text{silence}}) \cdot f_s / r \rceil$ embeddings,
where $d$ is the generation duration requested by the user, $d_{\text{silence}} {=} 6\,\text{s}$ is silence padding, $f_s {=} 44{,}100\,\text{Hz}$ is the sample rate, and $r {=} 4{,}096$ is the autoencoder downsampling ratio.
Only the first $L_{\text{eff}} = \lceil d \cdot f_s / r \rceil$ embeddings correspond to
the target audio content. The remaining $L - L_{\text{eff}}$ embeddings are silence padding.
Silence padding serves two purposes:
it prevents boundary artifacts that arise when the model produce an abrupt ending
exactly at the sequence edge, and it provides a fade-out buffer for the decoder.
After generation, the output can be trimmed to the requested $d$ seconds, discarding the padding region.

Our model does not require classifier-free guidance (CFG) at inference. Standard diffusion models rely on CFG to improve sample quality and text alignment, at the cost of two forward passes per denoising step (one conditional and one unconditional). In our pipeline, guidance is baked into the model through distillation warmup where the student is trained to match CFG-enhanced teacher trajectories, internalizing the quality boost that CFG provides. Adversarial post-training further refines text--audio alignment via the $\mathcal{L}_{\text{CLAP}}$ directly. As a result, our models do not rely on CFG during inference. This is a critical advantage for on-device deployment where memory and compute are constrained.

\section{Discussion}
\label{sec:discussion}

Stable Audio 3 is a family of text-prompted models for instrumental music and sound effects generation and editing, trained exclusively on licensed or creative commons data and designed to run fast and on consumer-grade hardware.

In the following sections, we discuss the results below:
\begin{itemize}
  \item State-of-the-art results for instrumental music and sound effects generation (Sections \ref{sec:eval_music} and \ref{sec:eval_sfx})
  \item Fast inference: less than 2s to generate up to 6m 20s on an H200 (Sections \ref{sec:eval_music} and \ref{sec:eval_sfx}).
  \item Robust variable-length audio generation (Sections \ref{sec:eval_varlen_music} and \ref{sec:eval_varlen_sfx}).
  \item Audio editing via inpainting, including single- and multi-segment edits and continuation (Section \ref{sec:eval_editing}).
  \item Adversarial post-training enables improved inference speed and sample quality (Section \ref{sec:eval_arc}).
  \item \texttt{small} and \texttt{medium} can run on consumer-grade GPUs and \texttt{small} on a MacBook Pro (Sections~\ref{subsec:vram_usage} and \ref{subsec:hw}).
\end{itemize}

\subsection{Methodology}
\label{sec:method}

We evaluate the Stable Audio 3 models against both open-weight and internal baselines using metrics and a subjective listening test. We compare a diverse set of systems spanning diffusion and autoregressive architectures:

\begin{itemize}

    \item \textbf{Stable Audio 2.5}~\cite{evans2024longform}: our internal latent diffusion baseline for (up to 190s) music generation. We use 8 sampling steps, CFG scale 6, and the DPM++ 3M SDE sampler.

    \item \textbf{Stable Audio Open}~\cite{evans2024stableaudioopen}: an open-weight latent diffusion model for (up to 47s) sound effects generation. We use 100 sampling steps, CFG scale 7, and the DPM++ 3M SDE sampler.

    \item \textbf{Stable Audio Open Small}~\cite{evans2024stableaudioopen}: a compact open-weight latent diffusion model optimized for efficient short-form (up to 11s) sound effects generation. We use 8 sampling steps and the PingPong sampler.

    \item \textbf{Stable Audio 3 `base model'}: our flow matching Stable Audio 3 variants (\texttt{small}, \texttt{medium}, \texttt{large}), without post-training. All models are evaluated using 50 sampling steps, CFG scale 7, and the Euler sampler.

    \item \textbf{Stable Audio 3 `post-trained'}: our Stable Audio 3 variants (\texttt{small}, \texttt{medium}, \texttt{large}) that are post-trained with distillation warmup and adversarial post-training. We use 8 sampling steps and PingPong sampler.

    \item \textbf{ACE-Step 1.5}~\cite{acestep2026}: an open-weight hybrid diffusion and autoregressive model for full-length song generation. We use the \texttt{acestep-5Hz-lm-4B} autoregressive model with the \texttt{acestep-v15-xl-turbo} backbone, since the authors found to provide a better quality--speed tradeoff than \texttt{acestep-v15-xl-sft}. Generation is performed using \textit{task\_type="text2music"}, \textit{lyrics="[Instrumental]"}, \textit{thinking=True}.

    \item \textbf{DiffRhythm 2}~\cite{jiang2026diffrhythm2}: an open-weight semi-autoregressive block flow-matching model for full-length song generation. Prompts are provided through the \texttt{style\_prompt} argument, while lyrics conditioning is disabled.

    \item \textbf{Woosh Flow}~\cite{hadjeres2026woosh}: an open-weight flow matching model for (up to 5s) sound effects generation. The model uses an adaptive-step ODE solver, resulting in 72 sampling steps on average.

    \item \textbf{Woosh DFlow}~\cite{hadjeres2026woosh}: an open-weight distilled version of Woosh Flow for (up to 5s) sound effects generation. We use 4 generation steps, following the official implementation.

    \item \textbf{TangoFlux}~\cite{hung2024tangoflux}: an open-weight flow matching model for (up to 30s) sound effects generation baseline. We use 50 sampling steps and CFG scale 4.5.

\end{itemize}

Note that different models generate audio at varying lengths. For a fair comparison, we evaluate all methods considering each model’s maximum supported duration, \textit{e.g.}, Woosh generates up to 5s or \texttt{small} generates up to 120s.

For models capable of vocal generation (ACE-Step 1.5 and DiffRhythm 2), evaluation is restricted to instrumental prompts to ensure fair comparison with Stable Audio 3, which focuses on instrumental music generation. We also considered JAM~\cite{jam2025}, HeartMuLa~\cite{heartmula2026}, and YuE~\cite{yue2025}. JAM was excluded because it relies on lyrics and reference audio conditioning instead of focusing on instrumental music generation. YuE and HeartMuLa were excluded because they rely on tags prompting and are designed for vocal music rather than instrumental music generation from text prompts.

All Stable Audio models trained on the AudioSparx dataset (Stable Audio 2.5 and 3 variants, excluding Stable Audio Open and Open Small) prepend prompts with \textit{"TrackType: Music, VocalType: Instrumental,"} for music generation and \textit{"TrackType: SFX,"} for sound effects generation. These prefixes indicate the target audio modality and significantly improve generation quality. We therefore recommend all Stable Audio 3 users to use those prefixes at inference time.

We report three metrics:

\begin{itemize}

    \item \textbf{Fr\'echet Audio Distance (FAD)}~\cite{kilgour2019fad,fadtk}: measures distributional similarity between generated and reference audio distributions. We compute FAD using embeddings from the LAION-CLAP audio encoder checkpoint \texttt{630k-audioset-best.pt}, which we found to provide the best perceptual correspondence. A low FAD implies that the generated audio is plausible and closely matches the reference audio.

    \item \textbf{CLAP score}: cosine similarity between text and audio embeddings from the same LAION-CLAP model above, measuring semantic alignment between prompts and generated audio. The higher the better.

    \item \textbf{Inference time}: wall-clock generation latency measured for fixed-length audio generation under standardized hardware and inference settings. Unless stated the contrary, inference time is measured on a H200.

\end{itemize}

We run our evaluation on two datasets:

\begin{itemize}

    \item \textbf{Song Describer Dataset (SDD)}~\cite{manco2023songdescriber}: a dataset of 120s-long music tracks paired with human-written captions. We exclude prompts containing vocals, speech, or sound effects, retaining only instrumental examples. We additionally remove incoherent or ambiguous prompts. This results in 424 music-caption pairs.

    \item \textbf{BBC Sound Effects Dataset}: a collection of recorded environmental and sound effects with text prompts.

\end{itemize}

For the BBC Sound Effects dataset, we first filter the dataset to samples with duration up to 120s. We select this cutoff as it is sufficient to cover both sound effects and environmental recordings, and matches the maximum generation length of \texttt{small}. Further, due to the duration limitations of several baselines, we construct multiple evaluation subsets:

\begin{itemize}

    \item {$\leq 120$s subset}: 10,491 audio-caption pairs. 

    \item {$\leq 30$s subset}: 5,406 audio-caption pairs. 

    \item {$\leq 10$s subset}: 1,537 audio-caption pairs. 

    \item {$\leq 5$s subset}: 393 audio-caption pairs. 

\end{itemize}

For all subsets above, generated audio durations match the duration of the corresponding reference. This variable-length setup enables fair comparison between models with different maximum generation durations while preserving realistic prompt-length distributions (\textit{e.g.}, shorter/longer durations for sound effects/environmental recordings).

We do not use the AudioCaps\cite{kim2019audiocaps} dataset for evaluation, as we found that its (reference) audio is bandwidth-limited. Instead, we use the BBC Sound Effects dataset, whose recordings are professionally produced and full-bandwidth.

We run a listening test using a Mean Opinion Score protocol with 14 participants, evaluating generated samples on:

\begin{itemize}

    \item \textbf{Overall quality (OVL)}: evaluates production quality, perceptual realism, and the absence of artifacts.

    \item \textbf{Text relevance (REL)}: evaluates how accurately the generated audio matches the conditioning text prompt.

    \item \textbf{Musicality (MUS)}: evaluates the capacity to articulate musically coherent melodies and harmonies.

\end{itemize}

For our inpainting music evaluations, we use the SDD with music of 120s long. For our inpainting evaluations with sound, we use BBC Sound Effects samples with durations between 30s and 120s (5088 audio-caption pairs), excluding shorter clips to ensure meaningful continuation and multi-region inpainting evaluation. We evaluate three settings:

\begin{itemize}

    \item \textbf{Single inpaint}: a randomly selected region corresponding to between 2\% and 20\% of the audio duration is masked and regenerated. The masked region is constrained to be at least 1s long.

    \item \textbf{Double inpaint}: two independent masked regions are generated using the same procedure as single inpainting. The two regions are constrained to be separated by at least 6s.

    \item \textbf{Continuation}: an initial segment is randomly selected (between 5s and 20\% of the audio) is preserved, and the remainder is regenerated until the end.

\end{itemize}

Identical randomly sampled inpainting regions are used across all compared models to ensure fair comparison.

In our inpainting evaluation these signals are available: the original audio (to be inpainted), the generated audio (with the inpainted part), and the prompt (used to guide inpainting). In our setup, the original audio and the prompt are paired and available through our evaluation datasets, and one of the models under evaluation generates part of the audio (via inpainting). Below, we use this terminology to describe the metrics we use to evaluate inpainting capabilities:

\begin{itemize}

    \item \textbf{FAD full} is computed between the (full) generated audio and the (full) original reference audio. This metric assesses how well the inpainted region integrates within the original audio, also taking into account the transitions between generated (inpainted) and original regions.

    \item \textbf{FAD inpaint} is computed only considering the generated (inpainted) regions and its corresponding reference (original) segments. This metric is useful to evaluate the generated (inpainted) part alone.

    \item \textbf{CLAP text-gen} measures the similarity between the text prompt and the generated (inpainted) region.

    \item \textbf{CLAP gen-orig} measures the similarity between the audio embeddings of the generated (inpainted) region and the original audio in this region.

\end{itemize}

For double-inpainting experiments, each inpainted region is treated as an independent evaluation, effectively doubling the number of evaluated inpaint segments. Yet, in this case, full-audio FAD metrics are computed only once per audio.

\subsection{Instrumental music generation}
\label{sec:eval_music}

We evaluate instrumental music generations on the SDD at two durations: 120s and 190s. The 120s setting corresponds to the maximum generation length of \texttt{small}, while the 190s setting evaluates longer-form generation and corresponds to the maximum generation length of Stable Audio 2.5. \texttt{small} is therefore excluded from the 190s evaluation.

\begin{table}[h]
\centering
\begin{tabular}{lccccccc}
\toprule
 &  &  &  &  &  &  & inference \\
 & length & FAD $\downarrow$ & CLAP $\uparrow$ & OVL $\uparrow$ & REL $\uparrow$ & MUS $\uparrow$ & time (s) $\downarrow$ \\
\midrule

DiffRhythm 2          & 120s & 0.293 & 0.158 & 3.05 $\pm$ 0.94 & 2.10 $\pm$ 1.29 & 2.60 $\pm$ 1.10 & 3.88 \\
ACE-Step 1.5 xl-turbo & 120s & 0.193 & 0.321 & 3.35 $\pm$ 1.09 & 3.30 $\pm$ 1.13 & 3.15 $\pm$ 1.31 & 6.23 \\
Stable Audio 2.5      & 120s & \textbf{{0.106}} & \textbf{{0.395}} & 3.90 $\pm$ 0.79 & \textbf{{4.30}} $\pm$ 0.66 & 3.70 $\pm$ 0.92 & 0.85 \\
\midrule

\texttt{small-music} & 120s & 0.145 & \textbf{{0.393}} & 3.20 $\pm$ 0.89 & 3.60 $\pm$ 0.94 & 3.15 $\pm$ 0.81 & 0.45 \\
\texttt{medium}      & 120s & \textbf{{0.107}} & \textbf{{0.390}} & \textbf{{4.20}} $\pm$ 0.89 & 4.25 $\pm$ 0.85 & 4.15 $\pm$ 0.93 & 0.78 \\
\texttt{large}       & 120s & \textbf{{0.101}} & \textbf{{0.393}} & 3.95 $\pm$ 0.89 & 3.80 $\pm$ 1.11 & \textbf{{4.30}} $\pm$ 0.73 & 0.81 \\
\bottomrule
\end{tabular}
\vspace{2mm}
\caption{Instrumental music generation results on the SDD dataset with 120s generations.}
\label{tab:music_results_120}
\end{table}

\begin{table}[h]
\centering
\begin{tabular}{lcccc}
\toprule
 & & & & inference \\
 & length & FAD $\downarrow$ & CLAP $\uparrow$ & time (s) $\downarrow$ \\
\midrule
DiffRhythm 2 & 190s & 0.307 & 0.088 & 5.85 \\
ACE-Step 1.5 xl-turbo & 190s & 0.184 & 0.273 & 9.21 \\
Stable Audio 2.5 & 190s & 0.128 & \textbf{{0.375}} & 0.85 \\
\midrule
\texttt{medium} & 190s & 0.116 & 0.362 & 0.88 \\
\texttt{large} & 190s & \textbf{{0.100}} & \textbf{{0.373}} & 0.93 \\
\bottomrule
\end{tabular}
\vspace{2mm}
\caption{Instrumental music generation results on SDD dataset with generations of 190s (\texttt{small} generates up to 120s).}
\label{tab:music_results_190}
\end{table}

Tables~\ref{tab:music_results_120} and~\ref{tab:music_results_190} report results across both settings. Overall, Stable Audio models achieve the strongest performance across metrics. Stable Audio 2.5 remains the best-performing baseline, while Stable Audio 3 \texttt{medium} and \texttt{large} substantially improve musicality. In contrast, the \texttt{small} variant performs noticeably worse than the larger models. However, despite its reduced size and the use of a lightweight autoencoder optimized for CPU inference, it remains competitive with open-weight baselines.
Further, Stable Audio models achieve substantially faster inference times than open-weight baselines, generating 190s audio in under one second. 
Finally, Stable Audio 3 models exhibit minimal degradation when increasing generation length from 120s to 190s. 
Overall, results highlight the efficiency advantages and improved musicality of Stable Audio 3, enabling high-quality long-form music generation with low latency.

\subsection{Sound effects generation} 
\label{sec:eval_sfx}

We evaluate sound effects generation using target durations of 5s, corresponding to the maximum supported duration of Woosh models. This restricted setting enables comparison against a broad set of recent open-weight sound effects generation systems. Additional evaluations at longer durations (up to 120s) are discussed in Section~\ref{sec:eval_varlen_sfx}.

\begin{table}[h]
\centering
\begin{tabular}{lcccccc}
\toprule
 &  &  &  &  & & inference \\
 & length & FAD $\downarrow$ & CLAP $\uparrow$ & OVL $\uparrow$ & REL $\uparrow$ & time (s) $\downarrow$ \\
\midrule

TangoFlux       & 5s & 0.760 & 0.179 & 2.35 $\pm$ 1.04 & 3.25 $\pm$ 1.37 & 1.90 \\
Woosh DFlow     & 5s & 0.619 & 0.228 & 3.10 $\pm$ 1.25 & 3.20 $\pm$ 1.64 & 0.06 \\
Woosh Flow      & 5s & 0.580 & 0.277 & 3.45 $\pm$ 1.19 & 3.80 $\pm$ 1.28 & 1.92 \\
SAO             & 5s & 0.501 & 0.263 & 2.95 $\pm$ 1.32 & 3.30 $\pm$ 1.30 & 12.30 \\
SAO-small       & 5s & 0.500 & 0.277 & 3.10 $\pm$ 1.12 & 3.55 $\pm$ 1.00 & 0.24 \\
\midrule

\texttt{small-sfx} & 5s & 0.395 & 0.351 & 3.35 $\pm$ 1.39 & 3.25 $\pm$ 1.45 & 0.41 \\
\texttt{medium}    & 5s & 0.369 & 0.369 & \textbf{\textit{3.65}} $\pm$ 1.14 & \textbf{\textit{3.95}} $\pm$ 1.23 & 0.60 \\
\texttt{large}     & 5s & \textbf{\textit{0.358}} & \textbf{\textit{0.370}} & 3.60 $\pm$ 0.94 & 3.85 $\pm$ 1.04 & 0.64 \\

\bottomrule
\end{tabular}
\vspace{2mm}
\caption{Sound effects generation results on the BBC Sound Effects Dataset with 5s generations.}
\label{tab:sfx_results}
\vspace{-2mm}
\end{table}

Table~\ref{tab:sfx_results} shows that Stable Audio 3 models consistently outperform all baselines. In particular, \texttt{large} and \texttt{medium} achieve the best overall performance while Woosh Flow remains a strong baseline. Interestingly, we observe a discrepancy between FAD and OVL scores for Woosh Flow. Although it attains competitive subjective OVL quality, it is often penalized by FAD due to producing band-limited signals. The results also highlight the efficiency of Stable Audio 3 models. While Woosh DFlow achieves extremely low inference latency, this comes at a quality cost. In contrast, Stable Audio 3 models maintain fast inference while obtaining state-of-the-art results for sound effects generation.

Unlike prior open-weight systems that specialize exclusively in either music or sound effects generation, \texttt{medium} and \texttt{large} are trained for both instrumental music and sound effects generation with a single model. Despite this shared training setup, they achieve state-of-the-art performance in both domains. Yet, due to the limited parameter budget of \texttt{small}, we instead train two specialized models: \texttt{small-music} for instrumental music generation and \texttt{small-sfx} for sound effects generation. This design choice enables improved quality despite tight compute and memory constraints.

\subsection{Variable-length instrumental music generation}
\label{sec:eval_varlen_music}

We evaluate Stable Audio models across multiple generation durations, ranging from 20s clips to 380s full-length generations (when possible). Note that \texttt{small-music} can generate up to 120s and Stable Audio 2.5 up to 190s. 

Stable Audio 2.5 was trained using fixed-length 190s sequences as in Figure \ref{fig:variable-length} (a). Shorter training examples were padded with silence to match the 190s maximum duration, enabling the model to learn variable-duration content within a fixed-length generation. Producing shorter clips, therefore, is not efficient as it requires running inference over the full sequence length, with most computation spent generating silence. Yet, running inference directly on shorter sequences than those used during training (as in Figure \ref{fig:variable-length} (b) despite being trained as (a)) could lead to degraded quality. We evaluate this behavior in the \textit{misused} Stable Audio 2.5 setting in Table \ref{tab:length_comparison_music_misuse}, where inference is performed directly at shorter durations instead of using the original fixed-length generation procedure. The results show that FAD and CLAP degrade, denoting that Stable Audio 2.5 does not generalize to perform \emph{efficient} variable-length inference.

\begin{table}[h]
\centering
\begin{tabular}{lcccc}
\toprule
 &  &  &  & inference \\
 & length & FAD $\downarrow$ & CLAP $\uparrow$ & time (s) $\downarrow$ \\
\midrule

Stable Audio 2.5 (misused) & 20s  & 0.731 & 0.285 & 0.41 \\
Stable Audio 2.5           & 20s & 0.149 & 0.389  & 0.85 \\ 
Stable Audio 2.5 (misused) & 120s & 0.201 & 0.382 & 0.79 \\
Stable Audio 2.5 & 120s & 0.106 & 0.395 & 0.85 \\
Stable Audio 2.5 & 190s & 0.128 & 0.375 & 0.85 \\

\bottomrule
\end{tabular}
\vspace{2mm}
\caption{Misusing Stable Audio 2.5 to perform efficient variable-length instrumental music generation without success.}
\label{tab:length_comparison_music_misuse}
\vspace{-2mm}
\end{table}

Stable Audio 3 models are explicitly designed for native variable-length generations. Our results in Table~\ref{tab:length_comparison_music} highlight the practical advantages of native variable-length generation. Stable Audio 3 inference cost scales naturally with output duration, enabling efficient short-form generation. 
Across durations, Stable Audio 3 models remain generally strong, with best performance typically observed at intermediate lengths (120--190s). At very short lengths (20s), we observe degradation in both FAD and CLAP, which we attribute to a mismatch between training and evaluation datasets, as most short training samples are loops (not full songs as in the evaluation set). At very long lengths (380s), performance also degrades, particularly the CLAP score. Informal listening suggests that the reduced prompt adherence is because long training examples are predominantly ambient or classical music. As a result, conditioning on long durations can bias the model toward generating ambient or classical music, omitting the provided text prompt.

\begin{table}[h]
\centering
\begin{tabular}{lcccc}
\toprule
 &  &  &  & inference \\
 & length & FAD $\downarrow$ & CLAP $\uparrow$ & time (s) $\downarrow$ \\
\midrule

\texttt{small-music}  & 20s  & 0.196      & 0.376     & 0.43 \\
\texttt{small-music}  & 120s & 0.145    & 0.393    & 0.45 \\

\midrule
\texttt{medium} & 20s  & 0.163 & 0.392 & 0.62 \\
\texttt{medium} & 120s & 0.107 & 0.390 & 0.78 \\
\texttt{medium} & 190s & 0.116 & 0.362 & 0.88 \\
\texttt{medium} & 380s & 0.156 & 0.275 & 1.31 \\

\midrule
\texttt{large}  & 20s  & 0.171 & 0.390 & 0.66 \\
\texttt{large}  & 120s & 0.101 & 0.393 & 0.81 \\
\texttt{large}  & 190s & 0.100 & 0.373 & 0.93 \\
\texttt{large}  & 380s & 0.131 & 0.277 & 1.80 \\

\bottomrule
\end{tabular}
\vspace{2mm}
\caption{Instrumental music generation results across different lengths.}
\label{tab:length_comparison_music}
\end{table}
\vspace{-5mm}
\begin{table}[h]
\centering
\begin{tabular}{lcccc}
\toprule
 &  &  &  & inference \\
 & length & FAD $\downarrow$ & CLAP $\uparrow$ & time (s) $\downarrow$ \\

\midrule
\texttt{small-sfx}    & 5s   & 0.395 & 0.35 & 0.41 \\
\texttt{small-sfx}    & 10s  & 0.367 & 0.348 & 0.44 \\
\texttt{small-sfx}    & 30s  & 0.335 & 0.343 & 0.46 \\
\texttt{small-sfx}    & 120s & 0.293 & 0.322 & 0.47\\

\midrule
\texttt{medium}   & 5s   & 0.369 & 0.369 & 0.60 \\
\texttt{medium}   & 10s  & 0.340 & 0.361 & 0.63 \\
\texttt{medium}   & 30s  & 0.297 & 0.358 & 0.65 \\
\texttt{medium}   & 120s & 0.266 & 0.321 & 0.77 \\

\midrule
\texttt{large}    & 5s   & 0.358 & 0.370 & 0.63 \\
\texttt{large}    & 10s  & 0.328 & 0.367 & 0.67 \\
\texttt{large}    & 30s  & 0.286 & 0.361 & 0.69 \\
\texttt{large}    & 120s & 0.259 & 0.322 & 0.81 \\

\midrule
SAO              & 5s   & 0.501 & 0.263 & 12.30 \\
SAO              & 10s  & 0.452 & 0.277 & 12.30 \\
SAO              & 30s  & 0.364 & 0.290 & 12.30 \\

\midrule
SAO-small        & 5s   & 0.500 & 0.277 & 0.24 \\
SAO-small        & 10s  & 0.456 & 0.280 & 0.24 \\

\midrule
TangoFlux        & 5s  & 0.760 & 0.179 & 1.90 \\
TangoFlux        & 10s   & 0.680 & 0.214 & 1.90 \\
TangoFlux        & 30s  & 0.595 & 0.206 & 1.90 \\

\midrule
Woosh Flow       & 5s   & 0.580 & 0.277 & 1.92 \\
Woosh DFlow      & 5s   & 0.619 & 0.228 & 0.06 \\

\bottomrule
\end{tabular}
\vspace{2mm}
\caption{Sound effects generation results across different lengths.}
\vspace{-5mm}
\label{tab:length_comparison}
\end{table}

\subsection{Variable-length sound effects generation}
\label{sec:eval_varlen_sfx}

Table~\ref{tab:length_comparison} evaluates the quality of sound effects generation and inference speed for varying output durations and models. Although \texttt{medium} and \texttt{large} support generation up to 380s, we restrict our evaluation to a maximum duration of 120s for simplicity and to easily compare against existing baselines. Across all evaluated durations, the proposed models consistently outperform the baselines while maintaining fast inference times.
An interesting trend is that FAD improves monotonically as generation length increases across all proposed variants. We hypothesize that this behavior is partly driven by the nature of longer-duration samples, which are predominantly composed of field recordings and ambient soundscapes with lower acoustic diversity and slower temporal variation. As a result, the generated audio exhibits lower distributional discrepancy with respect to the evaluation data, leading to improved FAD scores. In contrast, CLAP scores decrease for longer generations, likely reflecting the difficulty of maintaining semantic alignment with the text prompt over extended periods of time.

\subsection{Audio editing capabilities}
\label{sec:eval_editing}

In this section we evaluate the editing capabilities of our models on various tasks: inpainting (single- and double-region) and continuation, for both music (Table~\ref{tab:editing_music}) and sound effects
(Table~\ref{tab:editing_sfx}). 
Methodological details are explained in Section \ref{sec:method}.
Across both domains, \texttt{small} obtains worse FAD results than \texttt{medium} and \texttt{large}, which we attribute to its reduced model capacity and smaller (CPU-optimized) autoencoder.
Informal listening also reveals that \texttt{small} produces less smooth transitions, as reflected by the worse FAD full when compared to FAD inpaint. This discrepancy, however, is less pronounced in larger models, denoting that \texttt{medium} and \texttt{large} generate more coherent edits. Further, single and double inpaint numbers are close in both tables. This indicates that our models handle two independent masked regions as well as one.
Also, for the continuation setting FAD metrics are generally worse than inpainting. 
We attribute this to the stronger conditioning constraints in the inpainting setup, where the model is provided with a substantial amount of audio context. This conditioning effectively anchors inpaint edits to the reference distribution, reducing deviation and leading to lower FAD. In contrast, continuation requires extrapolating from a shorter context, which increases variability in long-range structure and results in worse FAD metrics.
Continuation results also show that FAD inpaint is worse than FAD full, the opposite of what we see in inpainting, because the generated region is unconstrained on one side and therefore drifts further from the reference distribution. The CLAP gen-orig metric is also worse, reflecting that without the surrounding context to anchor the generation, the continuations are less acoustically similar to the original even when prompt alignment (CLAP text-gen) remains high or improves.

\begin{table}[h]
\vspace{2mm}
\centering
\begin{tabular}{llccccc}
\toprule
 &  & FAD & FAD $\downarrow$ & CLAP $\uparrow$ & CLAP $\uparrow$ \\
 &  & full $\downarrow$ & inpaint & text-gen & gen-orig \\
\midrule

\multirow{4}{*}{Single Inpaint}
& \texttt{small}  & 0.100    & 0.086     & 0.271     & 0.833   \\
& \texttt{medium}  & 0.046 & 0.040 & 0.289 & 0.875 \\
& \texttt{large}  & 0.047 & 0.040 & 0.289 & 0.878 \\
\midrule

\multirow{4}{*}{Double Inpaint}
& \texttt{small}  & 0.100     & 0.083     & 0.274     & 0.834    \\
& \texttt{medium}  & 0.046 & 0.034 & 0.288 & 0.874 \\
& \texttt{large}  & 0.047 & 0.035 & 0.289 & 0.876 \\
\midrule

\multirow{4}{*}{Continuation}
& \texttt{small}  & 0.121     & 0.130     & 0.404     & 0.649     \\
& \texttt{medium}  & 0.074 & 0.084 & 0.400 & 0.653 \\
& \texttt{large}  & 0.071 & 0.078 & 0.403 & 0.667 \\

\bottomrule
\end{tabular}
\vspace{2mm}
\caption{Music editing results across models and tasks for inpainting and continuation settings.}
\label{tab:editing_music}
\vspace{-3mm}
\end{table}

\begin{table}[h]
\centering
\begin{tabular}{llccccc}
\toprule
 &  & FAD & FAD $\downarrow$ & CLAP $\uparrow$ & CLAP $\uparrow$ \\
 &  & full $\downarrow$ & inpaint & text-gen & gen-orig \\
\midrule

\multirow{3}{*}{Single Inpaint}
& \texttt{small} & 0.170 & 0.119 & 0.196 & 0.717 \\
& \texttt{medium} & 0.086 & 0.068 & 0.220 & 0.748 \\
& \texttt{large} & 0.084 & 0.068 & 0.220 & 0.752 \\
\midrule

\multirow{3}{*}{Double Inpaint}
& \texttt{small} & 0.171 & 0.119 & 0.195 & 0.714 \\
& \texttt{medium} & 0.089 & 0.070 & 0.216 & 0.742 \\
& \texttt{large} & 0.086 & 0.069 & 0.217 & 0.746 \\
\midrule

\multirow{3}{*}{Continuation}
& \texttt{small} & 0.208 & 0.214 & 0.351 & 0.600 \\
& \texttt{medium} & 0.172 & 0.191 & 0.332 & 0.570 \\
& \texttt{large} & 0.195 & 0.218 & 0.311 & 0.546 \\

\bottomrule
\end{tabular}
\vspace{2mm}
\caption{Sound effects editing results across models and tasks for inpainting and continuation settings.}
\label{tab:editing_sfx}
\end{table}

\subsection{Adversarial Post-Training discussion}
\label{sec:eval_arc}

Tables~\ref{tab:arc_rf_comparison_music} and ~\ref{tab:arc_rf_comparison_sfx} compare the pre-trained flow matching (base) models against models further trained using distillation warmup and adversarial post-training (post-trained). The base models require 50 sampling steps at inference time, resulting in substantially higher latency while also yielding inferior generation quality.
In contrast, post-trained models enable faster generation and can operate with as few as a single sampling step. However, directly generating audio latents from pure noise ($\epsilon$ $\to$ $\hat{x}_0$) in one step remains difficult, leading to degraded FAD and CLAP scores. For this reason, we choose 8-step ping-pong sampling in all our experiments. The iterative denoise-then-renoise schedule of ping-pong sampling allows the model to progressively refine its output, correcting errors from earlier steps while still leveraging the one-step denoising $x_t \to \hat{x}_0$ capability learned during distillation warmup and adversarial post-training.
Under this setting, post-trained models achieve a good balance between generation quality and efficiency, delivering improved FAD and CLAP scores while still requiring substantially lower inference time than the base models.

\begin{table}[h]
\centering
\begin{tabular}{llcccccc}
\toprule
 &  &  &  &  &  & inference \\
 &  & length & steps & FAD $\downarrow$ & CLAP $\uparrow$ & time (s) $\downarrow$ \\
\midrule

\texttt{small}  & base model  & 120s & 50 & 0.162 & 0.370 & 2.89 \\
\texttt{medium} & base model  & 120s & 50 & 0.143 & 0.352 & 3.87 \\
\texttt{large}  & base model & 120s & 50 & 0.116 & 0.355 & 3.90 \\

\midrule

\texttt{small}  & post-trained & 120s & 1  & 0.439 & 0.300 & 0.09 \\
\texttt{medium} & post-trained & 120s & 1  & 0.258 & 0.355 & 0.27 \\
\texttt{large}  & post-trained & 120s & 1  & 0.273 & 0.331 & 0.28 \\

\midrule

\texttt{small}  & post-trained & 120s & 8  & 0.145 & 0.393 & 0.45 \\
\texttt{medium} & post-trained & 120s & 8  & 0.107 & 0.390 & 0.78 \\
\texttt{large}  & post-trained & 120s & 8  & 0.101 & 0.393 & 0.81 \\

\bottomrule
\end{tabular}
\vspace{2mm}
\caption{Comparison of pre-trained and post-trained music models at various sampling steps.}
\label{tab:arc_rf_comparison_music}
\end{table}

\begin{table}[h]
\centering
\begin{tabular}{llcccccc}
\toprule
 &  &  &  &  &  & inference \\
 &  & length & steps & FAD $\downarrow$ & CLAP $\uparrow$ & time (s) $\downarrow$ \\
\midrule

\texttt{small}  & base model  & 120s & 50 & 0.336 & 0.284 & {2.89} \\
\texttt{medium} & base model  & 120s & 50 & 0.312 & 0.298 & 3.87 \\
\texttt{large}  & base model  & 120s & 50 & 0.282 & 0.299 & 3.90 \\

\midrule

\texttt{small}  & post-trained & 120s & 1  & 0.626 & 0.228 & 0.09 \\
\texttt{medium} & post-trained & 120s & 1  & 0.596 & 0.186 & 0.27 \\
\texttt{large}  & post-trained & 120s & 1  & 0.604 & 0.188 & 0.28 \\

\midrule

\texttt{small}  & post-trained & 120s & 8  & 0.293 & 0.322 & 0.45 \\
\texttt{medium} & post-trained & 120s & 8  & 0.266 & 0.321 & 0.78 \\
\texttt{large}  & post-trained & 120s & 8  & 0.259 & 0.322 & 0.81 \\

\bottomrule
\end{tabular}
\vspace{2mm}
\caption{Comparison of pre-trained and post-trained sound effects models at various sampling steps.}
\label{tab:arc_rf_comparison_sfx}
\end{table}

\subsection{VRAM Memory Usage}
\label{subsec:vram_usage}

In Table~\ref{tab:vram_usage} we report peak VRAM consumption across different models and generation durations. Peak memory usage increases with both model size and sequence length. 
\texttt{small} exhibits the lowest memory footprint, remaining below 2.5 GB even at 120s. \texttt{medium} and \texttt{large} require approximately 6.5 GB and 9.0 GB respectively at longer durations. 

These memory requirements are compatible with a wide range of modern consumer-grade GPUs. In particular, \texttt{small} can comfortably run on entry-level GPUs such as the RTX 3050 (typically 4--8 GB VRAM) and laptop GPUs with comparable memory capacity, while supporting generation lengths of up to 120s. Meanwhile, \texttt{medium} requires approximately 6.5 GB of VRAM for long-form audio generation and remains compatible with widely available consumer GPUs such as the RTX 3060 (12 GB VRAM), RTX 4060 (8 GB VRAM), and RTX 4070 (12 GB VRAM).

\begin{table}[h]
\centering
\begin{tabular}{lcc}
\toprule
 & Duration & Peak VRAM \\
\midrule

\texttt{small}  & 30s  & 1.89 GB \\
\texttt{small}  & 120s & 2.40 GB \\
\midrule

\texttt{medium} & 30s  & 5.49 GB \\
\texttt{medium} & 120s & 6.49 GB \\
\texttt{medium} & 380s & 6.52 GB \\
\midrule

\texttt{large}  & 30s  & 8.01 GB \\
\texttt{large}  & 120s & 9.01 GB \\
\texttt{large}  & 380s & 9.04 GB \\

\bottomrule
\end{tabular}
\vspace{2mm}
\caption{Peak VRAM usage across model sizes and generation durations.}
\label{tab:vram_usage}
\end{table}

\subsection{Inference Times Across Hardware Platforms}
\label{subsec:hw}

We compare inference times across multiple hardware platforms. All runs use a fixed configuration of 8 ping-pong sampling steps. We consider four different experiments: (i) MacBook Pro M4 with CPU-only; (ii) MacBook Pro M4 with CoreML acceleration employing CPU, GPU, and neural engine; (iii) NVIDIA H200 GPU with standard PyTorch execution; and (iv) NVIDIA H200 GPU with TensorRT acceleration. We report end-to-end generation latency across different generation durations and model scales. The results in Table~\ref{tab:hardware_inference_benchmark} show that CoreML significantly improves performance over CPU-only execution on the MacBook Pro M4. Second, on the H200 GPU, TensorRT further accelerates inference, reducing times by an order of magnitude in most configurations.

\paragraph{Implementation details}
CPU-only results on the MacBook Pro M4 are obtained through accelerating \texttt{small} with CoreML and accelerating SAME-S (decoder) with TFLite. The SAME-L decoders used by \texttt{medium} and \texttt{large} are accelerated in PyTorch with \texttt{torch.compile} instead of TensorRT. TensorRT does not support acceleration for the sliding-window attention used in SAME-L, whereas \texttt{torch.compile} can better exploit this setting.

\begin{table*}[h]
\centering
\begin{tabular}{lclcccc}
\toprule
 &  &  & \multicolumn{4}{c}{generation length} \\
Hardware & Acceleration & Model & 5s & 30s & 120s & 380s \\
\midrule

\multirow{2}{*}{MacBook Pro M4}
& CPU only  & \texttt{small} & 0.70s & 1.72s & 5.92s & --  \\ 
& CoreML  & \texttt{small}  & 0.23s & 0.63s & 3.09s & -- \\

\midrule

\multirow{6}{*}{H200}
& -- & \texttt{small}  & 0.41s & 0.46s & 0.45s & -- \\
& -- & \texttt{medium} & 0.60s & 0.65s & 0.78s & 1.31s \\
& -- & \texttt{large}  & 0.63s & 0.69s & 0.81s & 1.80s \\

& TensorRT & \texttt{small}  & 0.017s & 0.022s & 0.044s & -- \\
& TensorRT & \texttt{medium} & 0.02s & 0.05s & 0.13s & 0.43s \\
& TensorRT & \texttt{large}  & 0.03s & 0.07s & 0.19s & 0.63s  \\

\bottomrule
\end{tabular}
\vspace{2mm}
\caption{Inference times across hardware platforms, acceleration settings, model sizes, and generation lengths.}
\label{tab:hardware_inference_benchmark}
\end{table*}

\section{Conclusion}
\label{sec:conclusion}

Stable Audio 3 is a family of fast latent diffusion models (\texttt{small}, \texttt{medium}, \texttt{large}) for instrumental music and sound effects generation and editing. The models pair a semantic-acoustic autoencoder ($4096\times$ downsampling) with a diffusion transformer trained via flow matching, distillation warmup, and adversarial post-training. With only 8 ping-pong sampling steps at inference, they produce up to 6m~20s of stereo audio at 44.1\,kHz in under 2s on an H200 GPU. 
On instrumental music, Stable Audio 3 models improve musicality over prior work and outperform existing open-weight baselines. 
On sound effects, they likewise set a new state-of-the-art among open-weight systems. Our models natively support variable-length generation and inpainting-based editing, covering single- and multi-segment edits as well as continuation. Stable Audio 3 models are trained exclusively on licensed and Creative Commons data, and we release \texttt{small} and \texttt{medium} weights. \texttt{small} runs on a MacBook Pro M4 CPU, and \texttt{medium} fits on consumer GPUs with as little as 8\,GB of VRAM, putting both models within reach of typical research and creative workflows.

\bibliographystyle{unsrt}
\bibliography{sa3}

\end{document}